\documentclass[12pt]{article}
\usepackage[ruled]{algorithm}
\usepackage{graphicx,epsf,subfigure,pstricks,pst-node,psfrag,amsthm,amssymb,amsmath,fancyhdr,xfrac,booktabs,multirow,titling,titlesec,setspace,mathtools,nccmath,cuted,algorithmic,siunitx}
\usepackage[symbol]{footmisc}
\usepackage[font=small,labelfont=bf]{caption}

\usepackage[margin=1in]{geometry}
\usepackage{multirow}
\usepackage{rotating}
\usepackage{xcolor}
\usepackage{natbib}
\usepackage{subfig}
\newtheorem{prop}{Proposition}
\newtheorem{theorem}{Theorem}

\newtheorem{theoremC}{Theorem}
\newtheorem{assumptionC}{Assumption}
\newtheorem{lemmaC}{Lemma}
\newtheorem{definitionC}{Definition}

\DeclareMathOperator*{\argmin}{argmin} 
\begin{document} \sloppy
	\title{\vspace{-3ex}\textbf{\LARGE{Two-Dimensional Variable Selection and }}\vspace{-1ex} 
		\textbf{\LARGE{Its Applications in the Diagnostics of Product}}\vspace{-1ex} 
		\textbf{\LARGE{Quality Defects}}\\\vspace{1ex}
		\large{Cheoljoon Jeong and Xiaolei Fang}\vspace{-1ex}\\
		Edward P. Fitts Department of Industrial and Systems Engineering\vspace{-1ex}\\
		North Carolina State University\\\vspace{-1ex}}
	
	\date{\large{}\vspace{-10ex}}		
	\maketitle
	\textbf{Abstract.}	The root-cause diagnostics of product quality defects in multistage manufacturing processes often requires a joint identification of crucial stages and process variables. To meet this requirement, this paper proposes a novel penalized matrix regression methodology for two-dimensional variable selection. The method regresses a scalar response variable against a matrix-based predictor using a generalized linear model. The unknown regression coefficient matrix is decomposed as a product of two factor matrices. The rows of the first factor matrix and the columns of the second factor matrix are simultaneously penalized to inspire sparsity. To estimate the parameters, we develop a block coordinate proximal descent (BCPD) optimization algorithm, which cyclically solves two convex sub-optimization problems. We have proved that the BCPD algorithm always converges to a critical point with any initialization. In addition, we have also proved that each of the sub-optimization problems has a closed-form solution if the response variable follows a distribution whose (negative) log-likelihood function has a Lipschitz continuous gradient. A simulation study and a dataset from a real-world application are used to validate the effectiveness of the proposed method. \vspace{1ex} 	
	
	
\newpage
\section{Introduction}\label{s:intro}

\begin{figure}[htp!]
	\centering
	\hspace{-2mm}
	\includegraphics[width=0.8\textwidth]{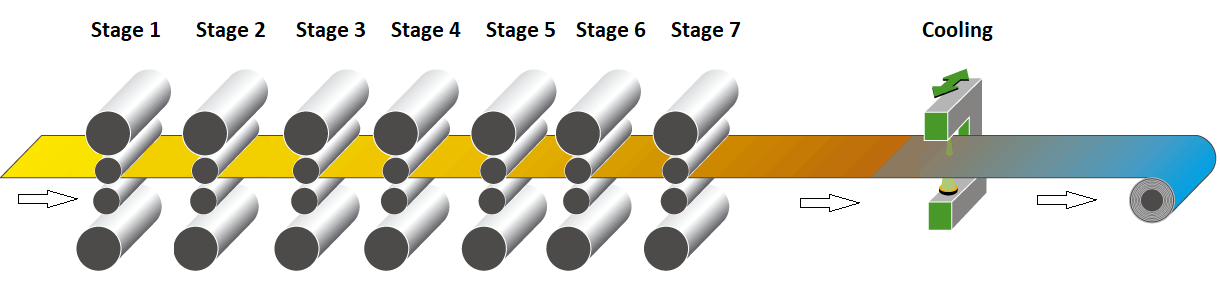}
	\caption{A hot strip mill with seven stands.}
	\label{fig:hotmill}
\end{figure}

\begin{figure}%
	\centering
	\subfigure[A product without a quality defect]{{\includegraphics[width=6cm]{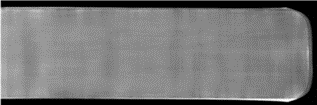} }}%
	\qquad
	\subfigure[A product with a quality defect]{{\includegraphics[width=5.6cm]{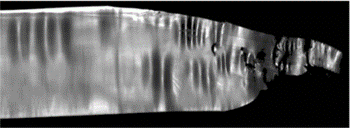} }}%
	\caption{Products with and without a quality defect from a hot rolling process \citep{Balmashnova2012}.}%
	\label{fig:qualitydefects}%
\end{figure}

Multistage manufacturing is a complex process that consists of multiple components, stations or stages to produce a product \citep{Shi2009}. For instance, Figure \ref{fig:hotmill} illustrates a seven-stage hot strip mill, the primary function of which is to roll heated steel slabs thinner and longer through seven successive rolling mill stands and then coil up the lengthened steel sheet for transport to the next process \citep{Ran2007}. The advancements in sensing technology and data acquisition systems have facilitated us to collect a massive amount of control and sensing data during the operation of such multistage processes. If modeled properly, these data could be very useful for system performance monitoring and diagnostics. System monitoring focuses on detecting defects/anomalies in real-time and diagnostics aims at identifying the root cause of the detected defects. One of the most common types of defects in multistage manufacturing is the product quality defect. For example, Figures \ref{fig:qualitydefects}(a) and \ref{fig:qualitydefects}(b) respectively demonstrate a product without and with a quality defect from the hot rolling process mentioned above. The diagnostics of the product quality defect in multistage manufacturing is of great importance since it helps locate the root cause of the defect and thus helps fix the abnormal process. This paper focuses on proposing a new method for the diagnostics of product quality defects in multistage manufacturing processes that all stages have a similar operation (such as the hot rolling process illustrated in Figure \ref{fig:hotmill}).

Multistage manufacturing processes typically involve a large number of control and sensing variables (referred to as ``process variables" hereafter) that potentially affect the quality of products. For example, the hot rolling process in Figure \ref{fig:hotmill} has more than 60 process variables, some of which are illustrated in Figure \ref{fig:pv}. A straightforward method to diagnose the root cause of product quality defects is LASSO \citep{Tibshirani1996}, which maps quality index (1 for a defected product and 0 otherwise) against process variables using logistic regression and penalizes the regression coefficients to inspire sparsity. Any process variable with a nonzero coefficient is considered to be responsible for the quality defect. However, one of the limitations of LASSO is that it cannot provide a structured solution that engineers can use to revise the control model to avoid further quality defects. To be specific, LASSO selects various process variables at various stages, which are difficult to be used to guide control model revision. In reality, due to the complexity of control theory, engineers are usually interested in identifying a few crucial process variables as well as stages that significantly affect product quality. For example, Table \ref{tb:x} shows a process variable matrix, the rows of which are process variables and columns represent stages. The cross markers represent the non-crucial process variables and stages for quality defects. For the multistage manufacturing process illustrated in Table \ref{tb:x}, engineers expect that process variables 1, 4 and 5 can be identified as crucial process variables, and stages 1 and 3  can be identified as crucial stages. To achieve this goal, one possible solution is to use group LASSO \citep{Yuan2007} to penalize the rows to identify crucial process variables first, and then remove the identified non-crucial rows and apply group LASSO again on the columns to select important stages. However, an obvious limitation of doing so is that the row selection accuracy is negatively affected by the non-crucial columns and vice versa. Consequently, the overall selection accuracy is compromised.

\renewcommand{\arraystretch}{1}

\begin{table}[htb!]
	\begin{center}
		\begin{tabular}{c|c|c|c|c|}	
			
			\multicolumn{1}{c}{} & \multicolumn{1}{c}{Stage} &  \multicolumn{1}{c}{Stage} & \multicolumn{1}{c}{Stage} & \multicolumn{1}{c}{Stage}\\
			\multicolumn{1}{c}{} & \multicolumn{1}{c}{1} &  \multicolumn{1}{c}{2} & \multicolumn{1}{c}{3} & \multicolumn{1}{c}{4}\\
			\cline{2-5}
			Process variable 1 &  & ${\times}$ &  &${\times}$  \\
			\cline{2-5}
			Process variable 2 &${\times}$ &${\times}$ &${\times}$  &${\times}$ \\
			\cline{2-5}
			Process variable 3 &${\times}$ &${\times}$ &${\times}$  &${\times}$ \\
			\cline{2-5}	
			Process variable 4 &  &$\times$ &  & $\times$ \\
			\cline{2-5}	
			Process variable 5 &  &$\times$ &  &$\times$ \\
			\cline{2-5}	
		\end{tabular}
	\end{center}
	\caption{An example process of the variable matrix (the cross markers represent the noncrucial process variables and stages).}
	\label{tb:x}
\end{table}

To address this challenge, this paper proposes a novel two-dimensional (2D) variable selection method that is capable of simultaneously identifying the crucial process variables and stages that are responsible for product quality defects. This is achieved by developing a penalized matrix regression model that regresses a product's quality index against its process variable matrix. Penalized matrix regression has been intensively studied in the existing articles \citep{Zhou2014, Zhao2014, Ding2018,Wang2017,Hung2010}. However, few of these existing models can conduct a 2D variable selection. The only exception is the structured LASSO \citep{Zhao2014}, which models the expectation of a normal distributed response variable ($y_i\in\mathbb{R}$) as the bilinear product of the explanatory matrix ($\mathbf{X}_i\in\mathbb{R}^{s\times t}$), i.e., $y_i=\boldsymbol{a}^\top\mathbf{X}_i\boldsymbol{b}+\epsilon_i$, where $\boldsymbol{a}\in\mathbb{R}^s$, $\boldsymbol{b}\in\mathbb{R}^t$, and $\epsilon_i\sim\mathcal{N}(0,\sigma^2)$. It achieves 2D variable selection by penalizing $\|\boldsymbol{a}\|_1\|\boldsymbol{b}\|_1$, where $\|\cdot\|_1$ is the $\ell_1$ norm. Numerical studies have indicated that structured LASSO is effective for 2D variable selection. However, structured LASSO assumes that the response variable follows a normal distribution, which is not necessary true for many applications. More importantly, Proposition 1 illustrates that the bilinear regression in structured LASSO is equivalent to the element-wise matrix regression $y_i = \big\langle\mathbf{B}, \mathbf{X}_i\big\rangle+\epsilon_i$ when $\mathbf{B}=\boldsymbol a\boldsymbol b^\top\in\mathbb{R}^{s\times t}$. This implies that structured LASSO assumes the rank of the regression coefficient matrix $\mathbf{B}$ is one, which significantly restricts its generality. Unlike structured LASSO, our proposed penalized matrix regression method assumes that the response variable follows a distribution from the exponential family (e.g., Bernoulli, binomial, normal, Poisson, gamma, exponential, etc.). In addition, we make no assumption on the rank of the coefficient matrix $\mathbf{B}$. Therefore, our method is more general than structured LASSO. To achieve 2D variable selection, we decompose the unknown regression coefficient matrix $\mathbf{B}$ as a product of two factor matrices $\mathbf{U}\in\mathbb{R}^{s\times r}$ and $\mathbf{V}\in\mathbb{R}^{r\times t}$, i.e., $\mathbf{B}=\mathbf{UV}$, where $r$ is the rank of the $\mathbf{B}$. We then simultaneously penalize the rows of factor matrix $\mathbf{U}$ and the columns of factor matrix $\mathbf{V}$ using adaptive group LASSO. The joint penalization on the two factor matrices results in both row-wise and column-wise sparsity of their product matrix (e.g., the estimated regression coefficient matrix $\hat{\mathbf{B}}$). As a result, any process variables (or stages) corresponding to the nonzero rows (or columns) of the regression coefficient matrix are considered as crucial process variables (or stages) that are responsible for product quality defects. To estimate the regression coefficient matrix, we develop a block coordinate proximal descent (BCPD) algorithm, which cyclically optimizes one of the two factor matrices until convergence. We will prove that the BCPD algorithm always converges to a critical point from any initialization point. In addition, we will also prove that each of the sub-optimization problems has a closed-form solution if the response variable (i.e., quality index) follows a distribution whose (negative) log-likelihood function has a Lipschitz continuous gradient.


The rest of this paper is organized as follows. In Section \ref{sec:method}, we introduce the 2D variable selection methodology. Section \ref{sec:optimization} presents the optimization algorithm. Sections \ref{sec:simulation} and \ref{sec:casestudy} validate the performance of our proposed 2D variable selection method using a numerical study and a real-world dataset, respectively. Finally, Section \ref{sec:conclusion} concludes.

\section{Two-Dimensional Variable Selection Methodology}\label{sec:method}

This paper proposes a 2D variable selection methodology that jointly identifies the crucial process variables and stages that are responsible for quality defects of a multistage manufacturing process. This is achieved by developing a penalized matrix regression method that regresses a product quality index against the process variable matrix using a generalized linear model. The unknown regression coefficient matrix is decomposed as the product of two factor matrices. The rows of the first factor matrix and the columns of the second factor matrix are simultaneously penalized using adaptive group LASSO, which results in an estimated coefficient matrix with sparse rows and columns. The variables (and stages) corresponding to the rows (and columns) of the coefficient matrix are considered to be crucial for the defect(s) of a product.

We assume that there exists a historical dataset for model training (aka. parameter estimation). The dataset consists of the quality index and process variables of $n$ products from the same multistage manufacturing process. We denote the quality index and process variable matrix of product $i$ as $y_i\in\mathbb{R}$ and $\mathbf{X}_i\in\mathbb{R}^{s\times t}$, respectively, where $s$ is the number of process variables and $t$ is the number of stages. As pointed out earlier, we assume that $y_i$ follows a distribution from the exponential family. As a result, its probability mass or density function can be expressed as follows \citep{McCullagh1983}:

\begin{equation}
\begin{aligned} 
\mathbb{P}(y_i|\theta,\phi) = \exp{\left\{\frac{y_i\theta-b(\theta)}{a(\phi)}+c(y_i,\phi)\right\}},
\end{aligned}\label{eq:link2}
\end{equation}

\noindent where $ \theta $ and $ \phi > 0 $ are parameters. $a(\cdot)$, $b(\cdot)$, and $c(\cdot)$ are known functions determined by the specific distribution in the exponential family. To construct the connection between the quality index and process variables, one possible way is to use the bilinear matrix regression proposed by \cite{Zhao2014}:

\begin{equation}
\begin{aligned} 
g(\mu_i)= \beta + \boldsymbol{a}^\top\mathbf{X}_i\boldsymbol{b},
\end{aligned}
\label{eq:link1}
\end{equation} 

\noindent where $\mu_i = \mathbb{E}(y_i|\textbf{X}_i)$ is the expectation of quality index. $g(\cdot)$ is the known link function that depends on the specific distribution that $y_i$ follows. $\beta$ is the intercept. $\boldsymbol{a}\in\mathbb{R}^s$ and $\boldsymbol{b}\in\mathbb{R}^t$. However, Proposition \ref{prop1} indicates that the model in Equation \eqref{eq:link1} is a special case of the following model using element-wise matrix regression

\begin{equation}
\begin{aligned} 
g(\mu_i)= \beta + \big\langle\mathbf{B}, \mathbf{X}_i\big\rangle,
\end{aligned}
\label{eq:link}
\end{equation} 

\noindent where $\mathbf{B}\in \mathbb{R}^{s\times t} $ is the regression coefficient matrix.  $\langle \cdot,\cdot \rangle$ is the matrix inner product operator, which is defined as $\big\langle\mathbf{B}, \mathbf{X}_i\big\rangle = \big\langle \text{vec}(\mathbf{B}), \text{vec}(\mathbf{X}_i) \big\rangle $ and $\text{vec}$ is the vectorization operator. 

\begin{prop}\label{prop1}
	Let $\boldsymbol{a}\in\mathbb{R}^{s}$, $\boldsymbol{b}\in\mathbb{R}^{t}$, and $\mathbf{B}=\boldsymbol{a}\boldsymbol{b}^\top$, then $\boldsymbol{a}^\top\mathbf{X}_i\boldsymbol{b}=\big\langle\mathbf{B}, \mathbf{X}_i\big\rangle$.  
\end{prop} 

The proof of the Proposition can be found in the appendix. Proposition 1 implies that Equation \eqref{eq:link1} assumes that the rank of the regression coefficient matrix $\mathbf{B}$ is one, which significantly restricts its generality. Therefore, in this paper, we will use Equation \eqref{eq:link} to construct the systematic part of the generalized linear model. The coefficient matrix $\mathbf{B}$ in Equation (\ref{eq:link}) can be estimated by using the maximum likelihood estimation (MLE) method, which maximizes the following log-likelihood function:
\begin{equation}
\begin{aligned} 
\ell\big(\mathbf{B},\beta\big)=\sum_{i=1}^n \frac{y_i\theta-b(\theta)}{a(\phi)}+\sum_{i=1}^n c(y_i,\phi),
\end{aligned}
\end{equation} 
where $ \theta $ is related to regression parameters $ (\mathbf{B},\beta) $. Let $ \mathcal{L}(\cdot)$ be the negative log-likelihood function, i.e., $ \mathcal{L}(\cdot) =-\ell(\cdot) $. MLE works by solving the following optimization problem:
\begin{equation}
\min_{\mathbf{B},\beta} \ 
\mathcal{L}(\mathbf{B},\beta).\ 
\end{equation}

To achieve both row-wise and column-wise selection, we decompose the unknown regression coefficient matrix as a product of two factor matrices: $ \mathbf{B}=\mathbf{U}\mathbf{V}$, where $\mathbf{U} \in \mathbb{R}^{s\times r} $, $ \mathbf{V} \in \mathbb{R}^{r\times t} $, and $r$ is the unknown rank that will be selected using model selection criteria such as AIC \citep{Akaike1974} or BIC \citep{Schwarz1978}. As a result, Equation \eqref{eq:link} can be expressed as follows:

\begin{equation}
\begin{aligned}
g(\mu_i) = \beta + \big\langle \mathbf{U}\mathbf{V},\mathbf{X}_i\big\rangle.
\end{aligned}
\end{equation} 

We then penalize the rows of $\mathbf{U}$ and columns of $\mathbf{V}$ respectively using adaptive group LASSO, which results in the following optimization criterion:

\begin{equation}
\begin{aligned}
\min_{\mathbf{U},\mathbf{V},\beta} \ 
\mathcal{L}(\mathbf{U},\mathbf{V},\beta) + \mathcal{R}(\mathbf{U},\mathbf{V}),
\end{aligned}
\label{eq:opt}
\end{equation}

\noindent where the regularization term

\begin{equation}\label{eq:reg}\nonumber
\begin{aligned}
\mathcal{R}(\mathbf{U},\mathbf{V}) = \lambda\gamma\left( \sum_{j=1}^s\frac{\Vert\mathbf{u}_j\Vert}{\Vert\hat{\mathbf{u}}_j\Vert}+\sum_{k=1}^t\frac{\Vert\mathbf{v}_k\Vert }{\Vert\hat{\mathbf{v}}_k\Vert}\right).
\end{aligned}
\end{equation}

\noindent Here, $ \lambda \ge 0 $ is the tuning parameter. $ \Vert\cdot\Vert $ is $ \ell_2 $ norm. $\mathbf{u}_j\in\mathbb{R}^r$ is the $j$th row of matrix $ \mathbf{U} $ and $\mathbf{v}_k\in\mathbb{R}^r$ is the $k$th column of matrix $ \mathbf{V}$. $ \hat{\mathbf{u}}_j $ and $ \hat{\mathbf{v}}_k $ are respectively the regular maximum likelihood estimates of $ \mathbf{u}_j $ and $\mathbf{v}_k$, which are constants and known. They are used to scale the penalized coefficients, which help address the estimation inefficiency and selection inconsistency challenges in group LASSO penalization (i.e., adaptive group LASSO, \cite{Wang2008}). $ \gamma = \sqrt{r} $, which is used to rescale the penalty with respect to the vector length of $ \mathbf{u}_j $ and $ \mathbf{v}_k $ ($ \mathbf{u}_j $ and $ \mathbf{v}_k $ have the same length $r$). 

The motivation behind this penalty term is that if a row of the factor matrix $ \mathbf{U} $ is penalized to be zeros, then the corresponding row of the estimated coefficient matrix $ \mathbf{B} $ will be zeros. Similarly, if a column of the factor matrix $ \mathbf{V} $ is penalized to be zeros, then the corresponding column of the estimated coefficient matrix $ \mathbf{B} $ will be zeros. Solving optimization criterion \eqref{eq:opt} provides a factor matrix $\hat{\mathbf{U}}$ with sparse rows and a factor matrix $\hat{\mathbf{V}}$ with sparse columns. Consequently, the estimated regression coefficient matrix $\hat{\mathbf{B}}=\hat{\mathbf{U}}\hat{\mathbf{V}}$ has both sparse rows and columns. The process variables corresponding to the nonzero rows of $\hat{\mathbf{B}}$ and the stages corresponding to the nonzero columns are identified as crucial ones for product quality defects.

\section{Optimization Algorithms}\label{sec:optimization}

In this section, we propose optimization algorithms to solve the optimization criterion \eqref{eq:opt}. In Section \ref{sec:sub:bcd}, we first propose a block coordinate descent (BCD) algorithm, which works by cyclically optimizing one block of parameters each time while keeping other blocks fixed. The subproblems (i.e., optimizing one block of parameters) of BCD are convex and thus can be solved using many existing convex optimization methods and packages. However, one of the limitations of BCD is that it is computationally intensive for applications with large datasets. To address this challenge, Section \ref{sec:sub:bcpd} proposes a block coordinate proximal descent (BCPD) algorithm that also cyclically optimizes one block of the parameters while keeping other blocks as a constant. In particular, we will prove that for these distributions whose (negative) log-likelihood functions possess Lipschitz continuous gradients, each subproblem of the BCPD algorithm has a closed-form solution, which significantly reduces the computation burden. For these distributions in the exponential family whose (negative) log-likelihood functions do not have Lipschitz continuous gradients, BCPD can also be used (without a closed-form solution for each subproblem).

\subsection{Block Coordinate Descent}\label{sec:sub:bcd}

As mentioned earlier, BCD works by cyclically optimizing one block of parameters each time while keeping other blocks fixed. Specifically, it iteratively optimizes $(\mathbf{U},\beta)$ with $\mathbf{V}$ fixed and then optimizes $(\mathbf{V}, \beta)$ with $\mathbf{U}$ fixed until convergence. Mathematically, this is achieved by cyclically solving the following two subproblems:

\begin{equation}
\begin{aligned}
(\mathbf{U}^k,\hat \beta^k)=\argmin_{\mathbf{U},\beta} \ 
\mathcal{L}(\mathbf{U},\mathbf{V}^{k-1},\beta) + \mathcal{R}(\mathbf{U},\mathbf{V}^{k-1}),
\end{aligned}
\label{eq:bcd1}
\end{equation}

\begin{equation}
\begin{aligned}
(\mathbf{V}^k,\beta^k)=\argmin_{\mathbf{V},\beta} \ 
\mathcal{L}(\mathbf{U}^k,\mathbf{V},\hat{\beta}^k) + \mathcal{R}(\mathbf{U}^k,\mathbf{V}).
\end{aligned}
\label{eq:bcd2}
\end{equation}

\noindent Algorithm \ref{ss:gg} summarizes the BCD algorithm.

\begin{algorithm}[H]
	\caption{Block Coordinate Descent} \label{ss:gg}
	\begin{algorithmic}[1]
		\STATE \textbf{Input:} $ \{\mathbf{X}_i,y_i\}_{i=1}^n $
		\STATE \textbf{Initialization:} Randomly choose $ \big(\mathbf{U}^0,\mathbf{V}^0,\beta^0\big) $
		\WHILE{convergence criterion not met}
		\STATE Compute $ \big(\mathbf{U}^k,\hat{\beta}^k\big) $ using \eqref{eq:bcd1}
		\STATE Compute $ \big(\mathbf{V}^k,\beta^k\big) $ using \eqref{eq:bcd2}
		\STATE Let $ k = k+1 $
		\ENDWHILE
	\end{algorithmic}

\end{algorithm}

Since both optimization subproblems \eqref{eq:bcd1} and \eqref{eq:bcd2} are convex, they can be solved using many existing algorithms and packages such as CVX and CVXQUAD \citep{Boyd2004}. However, existing convex optimization algorithms are computationally intensive and thus are not suitable for applications with large-size data. To address this challenge, we propose a BCPD algorithm in Section \ref{sec:sub:bcpd}.

\subsection{Block Coordinate Proximal Descent}\label{sec:sub:bcpd}
To accelerate the computation speed, we propose a BCPD algorithm that also cyclically optimizes one block of the parameters while keeping other blocks as a constant. To be specific, at iteration $k$, we solve the following subproblems:

\begin{equation}
\begin{aligned}\label{eq:gu}
\mathbf{U}^k = \argmin_{\mathbf{U}}\big\langle\nabla_\mathbf{U}\mathcal{L}\big(\mathbf{U}^{k-1},\mathbf{V}^{k-1},\beta^{k-1}\big), \mathbf{U}-\mathbf{U}^{k-1}\big\rangle +\frac{L_u^k}{2}\Vert
\mathbf{U}-\mathbf{U}^{k-1}\Vert^2_F + \mathcal{R}(\mathbf{U},\mathbf{V}^{k-1}),
\end{aligned}
\end{equation}
\begin{equation}\label{eq:gb1}
\begin{aligned}
\hat{\beta}^k = \argmin_\beta\big\langle\nabla_\beta\mathcal{L}\big(\mathbf{U}^{k-1},\mathbf{V}^{k-1},\beta^{k-1}\big), \beta-\beta^{k-1}\big\rangle  +\frac{L_u^k}{2}\big(\beta-\beta^{k-1}\big)^2, \\
\end{aligned}
\end{equation}
\begin{equation}\label{eq:gv}
\begin{aligned}\mathbf{V}^k = \argmin_{\mathbf{V}}\big\langle\nabla_\mathbf{V}\mathcal{L}\big(\mathbf{U}^{k},\mathbf{V}^{k-1},\hat{\beta}^k\big), \mathbf{V}-\mathbf{V}^{k-1}\big\rangle +\frac{L_v^k}{2}\Vert
\mathbf{V}-\mathbf{V}^{k-1}\Vert^2_F + \mathcal{R}(\mathbf{U}^k,\mathbf{V}),
\end{aligned}
\end{equation}
\begin{equation}\label{eq:gb2}
\begin{aligned}
\beta^k = \argmin_\beta\big\langle\nabla_\beta\mathcal{L}\big(\mathbf{U}^k,\mathbf{V}^{k-1},\hat{\beta}^k\big), \beta-\hat{\beta}^{k}\big\rangle  +\frac{L_v^k}{2}\big(\beta-\hat{\beta}^{k}\big)^2,
\end{aligned}
\end{equation}
where $ L_u^k $ and $ L_v^k $ are stepsize parameters, the values of which will be discussed later. For the distributions whose (negative) log-likelihood functions have Lipschitz continuous gradients (e.g., Bernoulli, binomial, and normal, etc.), Theorem \ref{th:closed-form} states that each subproblem has a closed-form solution. 

\begin{theorem}\label{th:closed-form}
 If the response variable $y_i$ follows a distribution whose (negative) log-likelihood function has a Lipschitz continuous gradient, subproblems \eqref{eq:gu}, \eqref{eq:gb1}, \eqref{eq:gv} and \eqref{eq:gb2} have the following closed-form solutions:
 \begin{equation}\label{eq:bcgdu}
 \begin{aligned}
 \mathbf{U}^k = \mathcal{S}_{\tau_u}\left(\mathbf{U}^{k-1}-\frac{1}{L_u^k}\nabla_\mathbf{U}\mathcal{L}\big(\mathbf{U}^{k-1},\mathbf{V}^{k-1},\beta^{k-1}\big)\right),
 \end{aligned}
 \end{equation}
 \begin{equation}\label{eq:bcgdb1}
 \begin{aligned}
 \hat{\beta}^k = \beta^{k-1} - \frac{1}{L_u^k}\nabla_\beta\mathcal{L}\big(\mathbf{U}^{k-1},\mathbf{V}^{k-1},\beta^{k-1}\big),
 \end{aligned}
 \end{equation}
 \begin{equation}\label{eq:bcgdv}
 \begin{aligned}
 \mathbf{V}^k = \mathcal{S}_{\tau_v}\left(\mathbf{V}^{k-1}-\frac{1}{L_v^k}\nabla_\mathbf{V}\mathcal{L}\big(\mathbf{U}^{k},\mathbf{V}^{k-1},\hat{\beta}^{k}\big)\right),
 \end{aligned}
 \end{equation}
 \begin{equation}\label{eq:bcgdb2}
 \begin{aligned} 
 \beta^k = \hat{\beta}^k -  \frac{1}{L_v^k}\nabla_\beta\mathcal{L}\big(\mathbf{U}^k,\mathbf{V}^{k-1},\hat{\beta}^k\big),
 \end{aligned}
 \end{equation}

\noindent where $ \tau_u=\lambda / (L_u^k\Vert\hat{\mathbf{u}}_j\Vert) $, $ \tau_v=\lambda / (L_v^k\Vert\hat{\mathbf{v}}_k\Vert) $, $ \mathcal{S}_{\tau_u}(\cdot) $ and $ \mathcal{S}_{\tau_v}(\cdot) $ are respectively the row-wise and column-wise soft-thresholding operator, which are defined below:
\begin{equation}
\begin{aligned}
(\mathcal{S}_{\tau_u}(\mathbf{u}))_j = 
\begin{cases}
\mathbf{u}_j-\tau_u\gamma\frac{\mathbf{u}_j}{\Vert\mathbf{u}_j\Vert}, & \mbox{if } \Vert\mathbf{u}_j\Vert > \tau_u\gamma \\
\mathbf{0}, & \mbox{if } \Vert\mathbf{u}_j\Vert \le \tau_u\gamma
\end{cases}
\end{aligned}
\end{equation}
\begin{equation}
\begin{aligned}
(\mathcal{S}_{\tau_v}(\mathbf{v}))_k = 
\begin{cases}
\mathbf{v}_k-\tau_v\gamma\frac{\mathbf{v}_k}{\Vert\mathbf{v}_k\Vert}, & \mbox{if } \Vert\mathbf{v}_k\Vert > \tau_v\gamma \\
\mathbf{0}, & \mbox{if } \Vert\mathbf{v}_k\Vert \le \tau_v\gamma
\end{cases}
\end{aligned}
\end{equation} 

\noindent for $ j = 1,2,...,s $ and $ k = 1,2,...,t $, where $ \gamma =\sqrt{r} $ and $r$ is the length of vectors $\mathbf{u}_j$ and $\mathbf{v}_k$ (they have the same length).

\end{theorem}

The proof of Theorem 1 can be found in the Appendix. We summarize the proposed BCPD algorithm in Algorithm \ref{ss:gg2} below. 

\begin{algorithm}
	\caption{Block Coordinate Proximal Descent} \label{ss:gg2}
	\begin{algorithmic}
		\STATE \textbf{Input:} $ \{\mathbf{X}_i,y_i\}_{i=1}^n $
		\STATE \textbf{Initialization:} Randomly choose $ \big(\mathbf{U}^0,\mathbf{V}^0,\beta^0\big) $
		\WHILE{convergence criterion not met}
		\STATE Compute $ \big(\mathbf{U}^k,\hat{\beta}^k\big) $ using \eqref{eq:bcgdu} and \eqref{eq:bcgdb1}
		\STATE Compute $ \big(\mathbf{V}^k,\beta^k\big) $ using \eqref{eq:bcgdv} and \eqref{eq:bcgdb2}
		\STATE Let $ k = k+1 $
		\ENDWHILE
	\end{algorithmic}
\end{algorithm}

To guarantee fast convergence, the stepsize parameters $ L_u^k $ and $ L_v^k $ are usually obtained from Lipschitz constants, which satisfy the following inequalities:

\begin{equation}
\begin{aligned}
\Vert\nabla_{(\mathbf{U},\beta)}\mathcal{L}\big(\mathbf{U},\mathbf{V}^{k-1},\beta\big)-\nabla_{(\mathbf{U},\beta)}\mathcal{L}\big(\tilde{\mathbf{U}},\mathbf{V}^{k-1},\tilde{\beta}\big)\Vert_F  \le L_u^k\Vert\big(\mathbf{U},\beta\big)-\big(\tilde{\mathbf{U}},\tilde{\beta}\big)\Vert_F, \\
\end{aligned}
\end{equation} 

\begin{equation}
\begin{aligned}
\Vert\nabla_{(\mathbf{V},\beta)}\mathcal{L}\big(\mathbf{U}^k,\mathbf{V},\beta\big)-\nabla_{(\mathbf{V},\beta)}\mathcal{L}\big(\mathbf{U}^k,\tilde{\mathbf{V}},\tilde{\beta}\big)\Vert_F \le L_v^k\Vert\big(\mathbf{V},\beta\big)-\big(\tilde{\mathbf{V}},\tilde{\beta}\big)\Vert_F,
\end{aligned}
\end{equation} 
 
\noindent where $ \Vert(\mathbf{U},\beta)\Vert_F = \sqrt{\Vert\mathbf{U}\Vert_F^2+\beta^2} $ and $ \Vert(\mathbf{V},\beta)\Vert_F = \sqrt{\Vert\mathbf{V}\Vert_F^2+\beta^2} $. For some of the distributions in the exponential family, we can easily derive the Lipschitz constants of the derivative of their (negative) log-likelihood function. For example, we have the following Lipschitz constants if the response variable of the generalized matrix regression is from the Bernoulli, binomial, or normal distribution (detailed derivation can be found in the Appendix):
\begin{equation}\label{eq:luconst}
\begin{aligned}
L_u = \sqrt{2}\ \sum_{i=1}^n\left(1+\Vert \mathbf{X}_i\mathbf{V}^\top\Vert_F\right)
\Vert 1+\mathbf{X}_i\mathbf{V}^\top\Vert_F, 
\end{aligned}
\end{equation}

\begin{equation}\label{eq:lvconst}
\begin{aligned}
L_v = \sqrt{2}\ \sum_{i=1}^n\left(1+\Vert \mathbf{U}^\top\mathbf{X}_i\Vert_F\right)
\Vert 1+\mathbf{U}^\top\mathbf{X}_i\Vert_F.
\end{aligned}
\end{equation}

Inspired by \citep{Shi2014}, the convergence criterion for the BCPD algorithm is defined using the relative objective function improvement and the relative change of the coefficient matrix:

\begin{equation}\label{eq:conv}
\begin{aligned}
q^k \equiv \max{\left\{\frac{\Vert\mathbf{B}^k-\mathbf{B}^{k-1}\Vert_F}{1+\Vert\mathbf{B}^{k-1}\Vert_F},\frac{|F^k-F^{k-1}|}{1+F^{k-1}}\right\}} \le \epsilon,
\end{aligned}
\end{equation}

\noindent where $\epsilon$ is a small number (e.g., $ 10^{-4} $), $ F^k = \mathcal{L}(\mathbf{U}^k,\mathbf{V}^k,\beta^k) + \mathcal{R}(\mathbf{U}^k,\mathbf{V}^k) $ is the objective function, and $ \mathbf{B}^k=\mathbf{U}^k\mathbf{V}^k $ is the estimated coefficient matrix at the $k$th step. Additionally, Theorem 2 indicates that the BCPD algorithm has a global convergence property, which implies that it converges to a critical point of optimization criterion \eqref{eq:opt} with any initialization.

\begin{theorem}[Global convergence]
	The sequence generated by the proposed BCPD algorithm converges to a critical point of the optimization criterion \eqref{eq:opt}.
\end{theorem}

\noindent The proof for Theorem 2 can be found in the Appendix. The closed-form solutions in Equations \eqref{eq:bcgdu}, \eqref{eq:bcgdb1}, \eqref{eq:bcgdv}, and \eqref{eq:bcgdb2} significantly reduce the computation time of the BCPD algorithm.

\section{Simulation Study}\label{sec:simulation}
In this section, we validate the performance of our 2D variable selection method using simulated datasets. 
\subsection{Data Generation}

We start with generating process variable matrix $\mathbf{X}_i$ by considering two correlation scenarios: (i) IID and (ii) row-wise correlation.
In the first scenario, all the entries of $\mathbf{X}_i$ are generated from an IID standard normal distribution. In the second scenario, we let the correlation between the row $j$ and row $k$ of matrix $\mathbf{X}_i$ be $0.5^{|j-k|}$ to mimic the correlation among process variables. 
For both scenarios, an IID noise matrix $\mathbf{E}_i$, each entry of which is from a standard normal distribution $\mathcal{N}(0,\sigma^2)$, is added to $\mathbf{X}_i$. Here, $\sigma$ is determined using a noise-to-signal ratio (NSR), which is defined as \textit{NSR} $={n\sigma}/\sum_{i=1}^n\|\mathbf{X}_i\|_F$, where $n$ is the sample size. We consider three levels of NSR, i.e.,``no noise'', ``low noise'', and ``high noise''. 
The NSRs for ``no noise'', ``low noise'', and ``high noise'' are respectively set as $0, 0.5$, and $1.0$.

Next, we generate the regression coefficient matrix. This is done by randomly generating two factor matrices $ \mathbf{U}\in \mathbb{R}^{s\times r}$ and $ \mathbf{V} \in\mathbb{R}^{r\times t}$ using MATLAB command \texttt{rand(s,r)*2-1} and \texttt{rand(r,t)*2-1}, where $ s=10, t=10, r=3 $. We then let the $1st, 3rd, 5th, 7th, 9th$ row of $ \mathbf{U} $ and the $2nd, 4th, 6th, 8th, 10th$ column of $ \mathbf{V} $ be zeros. The coefficient matrix is generated using $\mathbf{B}=\mathbf{U}\mathbf{V}$.

The product quality index (i.e., $y_i$) is generated using a logistic regression model $\log(\frac{y_i}{1-y_i})= \beta + \big\langle\mathbf{B},\mathbf{X}_i+\mathbf{E}_i\big\rangle$, where $\beta=0$. The generated data $\{\mathbf{X}_i,y_i\}_{i=1}^n$ are then used to validate the performance of our model. The stopping criteria is set as $ \epsilon=10^{-4}$ and the maximum number of iteration is $1,500$. We also consider three levels of data size $n=100, 200, 500$. 

The performance of our model is compared to three benchmarks. The first benchmark, designated as ``Row-Column'', first applies adaptive group LASSO to select the crucial rows of the process variable matrix $\mathbf{X}_i$. Next, the identified non-crucial rows are removed from $\mathbf{X}_i$ and adaptive group LASSO is then applied again to select the crucial columns. The second benchmarking model, which we refer to as ``Column-Row", is similar to the first benchmark except that it selects the crucial columns first and then identifies the crucial rows. The third benchmarking model is the structured LASSO \citep{Zhao2014}, which is similar to our proposed method except that it assumes the rank of the regression coefficient matrix is one (see Proposition 1 for details).

Since the optimization problem for our 2D variable selection method is nonconvex, the initial point is important for both the solution quality and convergence speed. Therefore, we propose the following heuristic method for parameter initialization. We first regress $ y_i $ against each entry of $ \mathbf{X}_i $ using logistic regression. The regression coefficients from all the entries are then used to construct a matrix $ \tilde{\mathbf{B}} $. Next, we apply singular value decomposition (SVD) on $ \tilde{\mathbf{B}} $, and set $ \mathbf{U}^0 $ to the first $ r $ (rank) left singular vectors and $ \mathbf{V}^0 $ to the first $ r $ right singular vectors of $ \tilde{\mathbf{B}} $. Various ranks are tested and the best rank is selected using AIC.

\subsection{Results and Analysis}
We apply our 2D variable selection method as well as the three benchmarks to the generated datasets. We compute the selection accuracy using the following equation:

\begin{equation}
\text{Accuracy} = \frac{\text{TP} + \text{TN}}{\text{Number of rows + Number of columns}},
\end{equation}

\noindent where ``TP" represents ``True Positive," which is the number of crucial rows and columns that are selected correctly. ``TN" represents ``True Negative," which is the number of non-crucial rows and columns that are removed correctly.

We repeat the whole simulation process for 100 times and report the mean selection accuracies and the corresponding standard deviations (SD) in Tables \ref{tb:a1} and \ref{tb:a2}.

\renewcommand{\arraystretch}{0.8}
\begin{table}[htb!]
	\begin{center}
		\footnotesize
		\begin{tabular}{c|c|c|cccc|c}
			\toprule
			Noise Level & \multirow{2}{*}{Method} & Sample & True & True & False & False& Accuracy \\ 
			(NSR) & & Size & Positive & Negative & Positive & Negative & (SD)\\
			\midrule 
			& \multirow{3}{*}{Proposed Method} & 100 & 99.2 & 94.1 & 5.9 & 0.8 & \textbf{96.7} (6.44)\\
			\multirow{12}{*}{No Noise} & & 200 & 99.2 & 99.4 & 0.6 & 0.8 & \textbf{99.3} (1.88) \\
			\multirow{12}{*}{(0)} & & 500 & 100.0 & 100.0 & 0.0 & 0.0 & \textbf{100.0} (0.00)\\
			\cmidrule(r){2-8}
			& \multirow{3}{*}{Benchmark I} & 100 & 95.7 & 90.1 & 9.9 & 4.3 & 92.9 (6.44) \\
			& & 200 & 98.6 & 99.5 & 0.5 & 1.4 & 99.1 (2.90) \\
			& & 500 & 100.0 & 100.0 & 0.0 & 0.0 & 100.0 (0.00)\\
			\cmidrule(r){2-8}
			& \multirow{3}{*}{Benchmark II} & 100 & 97.8 & 88.7 & 11.3 & 2.2 & 93.3 (4.46)\\
			& & 200 & 98.7 & 97.8 & 2.2 & 1.3 & 98.3 (2.40)\\
			& & 500 & 99.9 & 100.0 & 0.0 & 0.1 & 100.0 (0.50)\\
			\cmidrule(r){2-8}
			& \multirow{3}{*}{Benchmark III} & 100 & 99.0 & 94.0 & 6.0 & 1.0 & 96.5 (4.12)\\
			& & 200 & 99.0 & 98.0 & 2.0 & 1.0 & 98.5 (2.42)\\
			& & 500 & 100.0 & 100.0 & 0.0 & 0.0 & 100.0 (0.00)\\
			\midrule
			& \multirow{3}{*}{Proposed Method} & 100 & 98.2 & 93.4 & 6.6 & 1.8 & \textbf{95.8} (3.23)\\
			\multirow{12}{*}{Low Noise} & & 200 & 98.8 & 99.4 & 0.6 & 1.2 & \textbf{99.1} (2.29)\\
			\multirow{12}{*}{(0.5)} & & 500 & 100.0 & 100.0 & 0.0 & 0.0 & \textbf{100.0} (0.00)\\
			\cmidrule(r){2-8}
			& \multirow{3}{*}{Benchmark I} & 100 & 93.7 & 89.5 & 10.5 & 6.3 & 91.6 (6.85) \\
			& & 200 & 98.2 & 99.6 & 0.4 & 1.8 & 98.9 (3.06) \\
			& & 500 & 100.0 & 100.0 & 0.0 & 0.0 & 100.0 (0.00)\\
			\cmidrule(r){2-8}
			& \multirow{3}{*}{Benchmark II} & 100 & 96.9 & 88.4 & 11.6 & 3.1 & 92.7 (4.74) \\
			& & 200 & 98.1 & 97.6 & 2.4 & 1.9 & 97.9 (2.59) \\
			& & 500 & 99.8 & 100.0 & 0.0 & 0.2 & 99.9 (0.70) \\
			\cmidrule(r){2-8}
			& \multirow{3}{*}{Benchmark III} & 100 & 98.0 & 90.0 & 10.0 & 2.0 & 94.0 (3.94) \\
			& & 200 & 97.0 & 97.0 & 3.0 & 3.0 & 97.0 (3.50) \\
			& & 500 & 100.0 & 100.0 & 0.0 & 0.0 & 100.0 (0.00) \\
			\midrule
			& \multirow{3}{*}{Proposed Method} & 100 & 96.3 & 93.2 & 6.8 & 3.7 & \textbf{94.8} (3.72)\\
			\multirow{12}{*}{High Noise} & & 200 & 98.3 & 99.2 & 0.8 & 1.7 & \textbf{98.8} (2.79) \\
			\multirow{12}{*}{(1.0)} & & 500 & 99.6 & 100.0 & 0.0 & 0.4 & \textbf{99.8} (0.98)\\
			\cmidrule(r){2-8}
			& \multirow{3}{*}{Benchmark I} & 100 & 89.0 & 89.5 & 10.5 & 11.0 & 89.3 (7.50)\\
			& & 200 & 96.5 & 99.5 & 0.5 & 3.5 & 98.0 (4.14) \\
			& & 500 & 98.1 & 100.0 & 0.0 & 1.9 & 99.1 (4.42) \\
			\cmidrule(r){2-8}
			& \multirow{3}{*}{Benchmark II} & 100 & 92.1 & 89.3 & 10.7 & 7.9 & 90.7 (8.04)\\
			& & 200 & 96.0 & 98.2 & 1.8 & 4.0 & 97.1 (3.71) \\
			& & 500 & 96.3 & 100.0 & 0.0 & 3.7 & 98.2 (4.59) \\
			\cmidrule(r){2-8}
			& \multirow{3}{*}{Benchmark III} & 100 & 94.0 & 91.0 & 9.0 & 6.0 & 92.5 (4.86)\\
			& & 200 & 87.0 & 98.0 & 2.0 & 13.0 & 92.5 (15.14) \\
			& & 500 & 84.0 & 100.0 & 0.0 & 16.0 & 92.0 (15.67) \\
			\bottomrule 
		\end{tabular}
	\end{center}
	\vspace{-2ex}
	\caption{Average true positive (\%), true negative (\%), false positive (\%), false negative (\%), selection accuracy (\%) for $\mathbf{X}_i$ with IID entries.}
	\label{tb:a1}
\end{table}
 
\renewcommand{\arraystretch}{0.8}
\begin{table}[htb!]
	\begin{center}
		\footnotesize
		\begin{tabular}{c|c|c|cccc|c}
			\toprule
			Noise Level & \multirow{2}{*}{Method} & Sample & True & True & False & False & Accuracy \\
			(NSR) & & Size & Positive & Negative & Positive & Negative & (SD) \\
			\midrule
			& \multirow{3}{*}{Proposed Method} & 100 & 97.2 & 93.4 & 6.6 & 2.8 & \textbf{95.3} (3.75) \\
			\multirow{12}{*}{No Noise} & & 200 & 99.4 & 99.5 & 0.5 & 0.6 & \textbf{99.5} (1.57)\\
			\multirow{12}{*}{(0)} & & 500 & 100.0 & 100.0 & 0.0 & 0.0 & \textbf{100.0} (0.0)\\
			\cmidrule(r){2-8}
			& \multirow{3}{*}{Benchmark I} & 100 & 89.1 & 84.3 & 15.7 & 10.9 & 86.7 (7.11)\\
			& & 200 & 85.9 & 99.0 & 1.0 & 14.1 & 92.5 (7.54)\\
			& & 500 & 87.2 & 100.0 & 0.0 & 12.8 & 93.6 (4.43) \\
			\cmidrule(r){2-8}
			& \multirow{3}{*}{Benchmark II} & 100 & 94.6 & 87.1 & 12.9 & 5.4 & 90.9 (5.99) \\
			& & 200 & 96.6 & 98.0 & 2.0 & 3.4 & 97.3 (3.79) \\
			& & 500 & 99.8 & 100.0 & 0.0 & 0.2 & 99.9 (1.00)\\
			\cmidrule(r){2-8}
			& \multirow{3}{*}{Benchmark III} & 100 & 96.0 & 92.0 & 8.0 & 4.0 & 94.0 (6.58) \\
			& & 200 & 97.0 & 99.0 & 1.0 & 3.0 & 98.0 (3.50) \\
			& & 500 & 100.0 & 100.0 & 0.0 & 0.0 & 100.0 (0.00)\\
			\midrule
			& \multirow{3}{*}{Proposed Method} & 100 & 96.1 & 93.3 & 6.7 & 3.9 & \textbf{94.7} (4.65)\\
			\multirow{12}{*}{Low Noise} & & 200 & 98.4 & 99.5 & 0.5 & 1.6 & \textbf{99.0} (2.39)\\
			\multirow{12}{*}{(0.5)} & & 500 & 100.0 & 100.0 & 0.0 & 0.0 & \textbf{100.0} (0.00)\\
			\cmidrule(r){2-8}
			& \multirow{3}{*}{Benchmark I} & 100 & 86.7 & 85.0 & 15.0 & 13.3 & 85.9 (7.95) \\
			& & 200 & 85.0 & 98.7 & 1.3 & 15.0 & 91.9 (7.74) \\
			& & 500 & 87.3 & 100.0 & 0.0 & 12.7 & 93.7 (4.43) \\
			\cmidrule(r){2-8}
			& \multirow{3}{*}{Benchmark II} & 100 & 93.4 & 87.4 & 12.6 & 6.6 & 90.4 (5.76) \\
			& & 200 & 95.8 & 98.3 & 1.7 & 4.2 & 97.1 (4.98) \\
			& & 500 & 99.8 & 100.0 & 0.0 & 0.2 & 99.9 (0.70) \\
			\cmidrule(r){2-8}
			& \multirow{3}{*}{Benchmark III} & 100 & 96.0 & 90.0 & 10.0 & 4.0 & 93.0 (5.37) \\
			& & 200 & 98.0 & 99.0 & 1.0 & 2.0 & 98.5 (2.42) \\
			& & 500 & 100.0 & 100.0 & 0.0 & 0.0 & 100.0 (0.00)\\
			\midrule
			& \multirow{3}{*}{Proposed Method} & 100 & 92.0 & 92.3 & 7.7 & 8.0 & \textbf{92.2} (6.83) \\
			\multirow{12}{*}{High Noise} & & 200 & 97.2 & 99.2 & 0.8 & 2.8 & \textbf{98.2} (3.37)\\
			\multirow{12}{*}{(1.0)} & & 500 & 99.3 & 100.0 & 0.0 & 0.7 & \textbf{99.7} (1.28)\\
			\cmidrule(r){2-8}
			& \multirow{3}{*}{Benchmark I} & 100 & 78.5 & 87.4 & 12.6 & 21.5 & 83.0 (10.66)\\
			& & 200 & 82.2 & 99.5 & 0.5 & 17.8 & 90.9 (8.50) \\
			& & 500 & 86.9 & 100.0 & 0.0 & 13.1 & 93.5 (5.11) \\
			\cmidrule(r){2-8}
			& \multirow{3}{*}{Benchmark II} & 100 & 87.4 & 89.7 & 10.3 & 12.6 & 88.6 (7.73) \\
			& & 200 & 91.0 & 99.0 & 1.0 & 9.0 & 95.0 (7.07)\\
			& & 500 & 97.9 & 100.0 & 0.0 & 2.1 & 99.0 (3.12) \\
			\cmidrule(r){2-8}
			& \multirow{3}{*}{Benchmark III} & 100 & 88.0 & 90.0 & 10.0 & 12.0 & 89.0 (7.75) \\
			& & 200 & 97.0 & 99.0 & 1.0 & 3.0 & 98.0 (2.58) \\
			& & 500 & 99.8 & 100.0 & 0.0 & 2.0 & 99.9 (2.11)\\
			\bottomrule
		\end{tabular}
	\end{center}
	\vspace{-2ex}
	\caption{Average true positive (\%), true negative (\%), false positive (\%), false negative (\%), selection accuracy (\%) for $\mathbf{X}_i$ with row-wise correlated entries.}
	\label{tb:a2}
\end{table}

Table \ref{tb:a1} summarizes the variable selection accuracies (precisions) for the process variable matrices that are generated from an IID standard normal distribution. Table \ref{tb:a1} indicates that our proposed model and the three benchmarks usually achieve higher selection accuracies (precisions) with larger sample size. For example, when NSR is zero, the selection accuracies (precisions) of our model are $96.7 (6.44)$, $99.3(1.88)$, and $100.0(0.00)$ for sample size $100$, $200$, and $500$, respectively. Another example is that the selection accuracies (precisions) of benchmark I with a low NSR are respectively $91.6 (6.85)$, $98.9(3.06)$, and $100.0 (0.00)$ for sample size $100$, $200$, and $500$. This is reasonable since more samples result in a more accurate model estimation and thus a higher selection accuracy (precision). Table \ref{tb:a1} also illustrates that our model and the benchmarks usually have a lower selection accuracy (precision) when the NSR is higher. For example, when the sample size is $200$, the selection accuracy (precision) of our model is $99.3 (1.88)$ for ``No Noise", $99.1 (2.29)$ for ``Low Noise", and $98.8 (2.79)$ for ``High Noise". Similar phenomenon can also be observed for benchmarks I, II, and III. This is also justifiable since signals with a higher level of noise compromise the accuracy of model estimation and thus the accuracy of variable selection. Another observation from Table \ref{tb:a1} is that our proposed model almost always accomplishes a higher selection accuracy (precision) than the three benchmarks at all levels of NSR and all levels of sample size. For example, the selection accuracies (precisions) of our method and the three benchmarking models are respectively $99.3(1.88)$, $99.1(2.90)$, $98.3(2.40)$, and $98.5(2.42)$ when the sample size is 200 and the NSR is zero. Similarly, they are respectively $95.8(3.23)$, $91.6(6.85)$, $92.7(4.74)$, and $94.0(3.94)$ when the sample size is 100 and the NSR is low. We believe the reason why our proposed method outperforms benchmarks I and II is that our method simultaneously identifies the crucial rows and columns but the benchmark models I and II select crucial rows and then columns (or columns and then rows) sequentially. For example, benchmark I selects the crucial rows (first step) and then identifies the important columns (second step). One drawback of doing so is that the non-crucial columns may have a significant negative effect on the accuracy of row selection in the first step and the non-accurately selected rows will in turn compromise the column selection accuracy in the second step. The same limitation also applies to benchmark II. We believe the reason why our proposed method has superiority over benchmark III (structured LASSO) is because its flexibility in modeling the rank of the regression coefficient matrix. As pointed out earlier, our method has no assumption on the rank of the regression coefficient matrix, whereas structured LASSO assumes the rank is one. In this numerical study, the true rank of the coefficient matrix is 3, which violates the assumption of benchmark III.

Table \ref{tb:a2} reports the variable selection accuracies (precisions) for process variable matrices with row-correlated entries. Similar to Table \ref{tb:a1}, we observe from Table \ref{tb:a2} that our proposed 2D variable selection method achieves a higher variable selection accuracy (precision) than the three benchmarks at all levels of NSR and all levels of sample size. This again illustrates the superiority of our proposed method. Table \ref{tb:a2} also suggests that the selection accuracy (precision) of benchmark II is better than that of benchmark I at all levels of NSR and all levels of sample size. For example, when there is no noise and the sample size is $200$, the selection accuracy (precision) is $92.5(7.54)$ for benchmark I and $97.3(3.79)$ for benchmark II. When the NSR is low and the sample size is $500$, the selection accuracies (precisions) for benchmarks I and II are respectively $93.7(4.43)$ and $99.9(0.70)$. Similarly, when the NSR is high and the sample size is $100$, they are $83.0(10.66)$ and $88.6(7.73)$ respectively. We believe this is due to the fact that the process variable matrix is row-wise correlated. Specifically, since benchmark I selects the crucial rows first, the row-wise correlation compromises its selection accuracy. Unlike benchmark I, benchmark II identifies the important columns first, which is not significantly affected by the row-wise correlation of the process variable matrix.


\section{Case Study}\label{sec:casestudy}

In this section, we apply our proposed 2D variable selection method to a dataset from a real-world application and compare this method with three benchmarking methods. The dataset is from the hot strip mill illustrated in Figure \ref{fig:hotmill}. The primary function of the hot strip mill is to reheat semi-finished steel slabs nearly to their melting point, and then roll them thinner and longer through 7 successive rolling mill stands driven by motors, and finally coil up the lengthened steel sheet for transport to the next process. 

\begin{figure}[htp!]
	\centering
	\qquad \qquad 
	\subfigure[Measurement points for width]{\raisebox{+2cm}{\includegraphics[width=5cm]{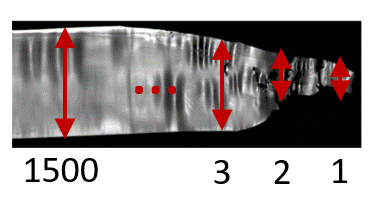} }}%
	\qquad \quad
	\subfigure[Difference between desired and measured width]{{\includegraphics[width=8cm]{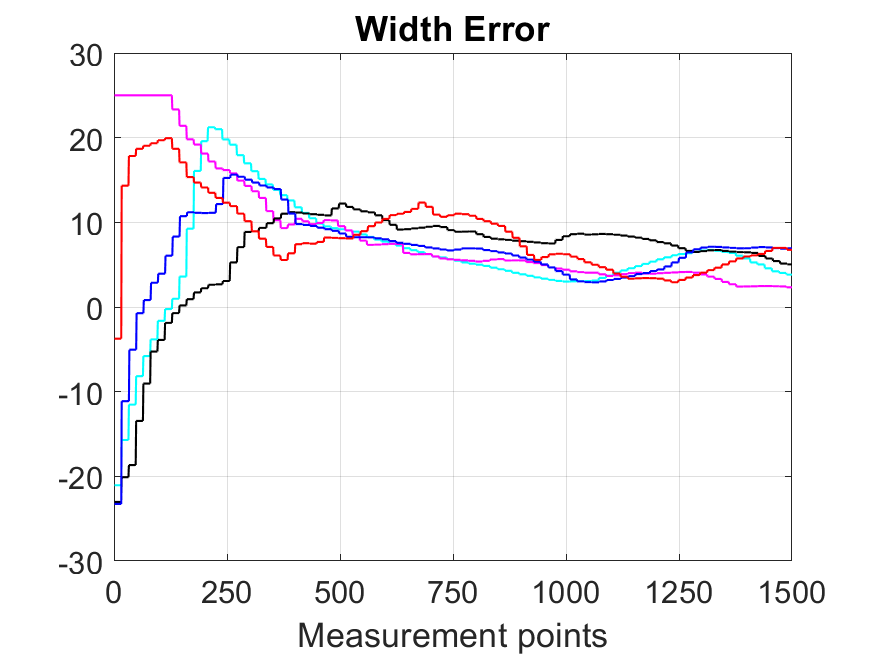} }}%
	\caption{Measurement points of steel strip products and corresponding width error.}%
	\label{fig:qualitydefectsmeasurement}%
\end{figure}

The dataset consists of $490$ strip steel products made using the same hot strip mill. Among the $490$ samples, $264$ of them are good quality products and the remaining $226$ are products with defects. The product quality is defined based on the percentage of width that is smaller than a predefined width threshold. Specifically, the width of each product (i.e., steel strip) was measured at $1,500$ locations uniformly distributed along the head of the strip. Figure \ref{fig:qualitydefectsmeasurement}(a) illustrates the measurement points and Figure \ref{fig:qualitydefectsmeasurement}(b) indicates the measured width error of five strip products. Any measurement point whose width is smaller than a predefined width threshold is considered as a defect point. The quality index of each product is then constructed by dividing the number of its width defect points by the number of measurement points (i.e., $1,500$). Therefore, the range of the quality index is $[0,1]$. 

\begin{figure}[htp!]
	\vspace{1mm}
	\centering
	\subfigure[Process variable profiles at Stage 5]{{\includegraphics[width=7.6cm]{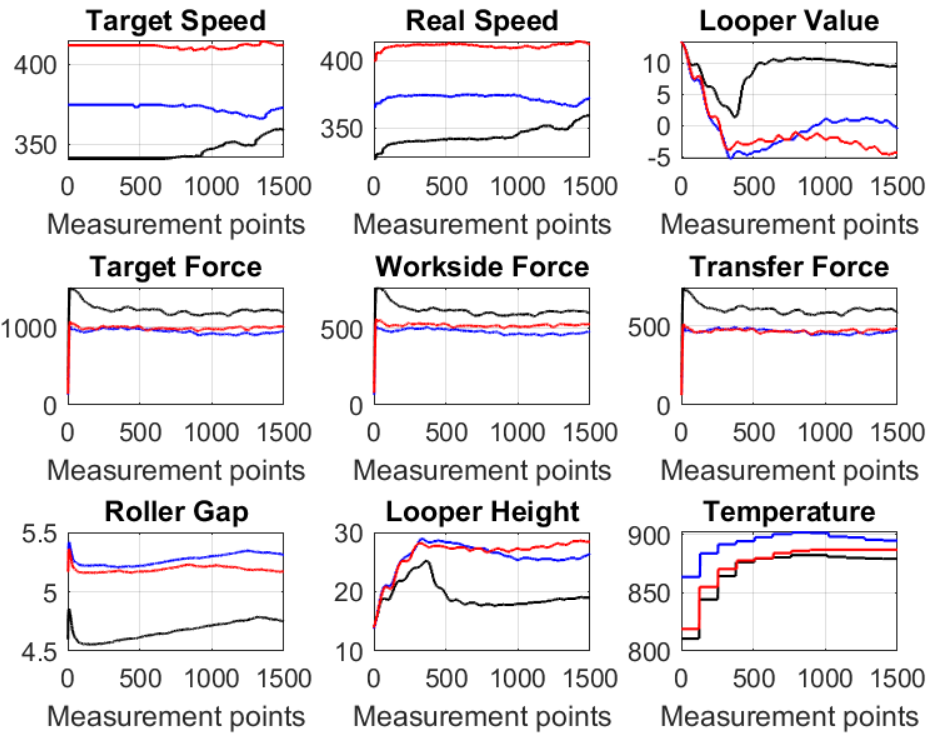} }}%
	\qquad
	\subfigure[Average process variable measurements at all the stages]{{\includegraphics[width=7.3cm]{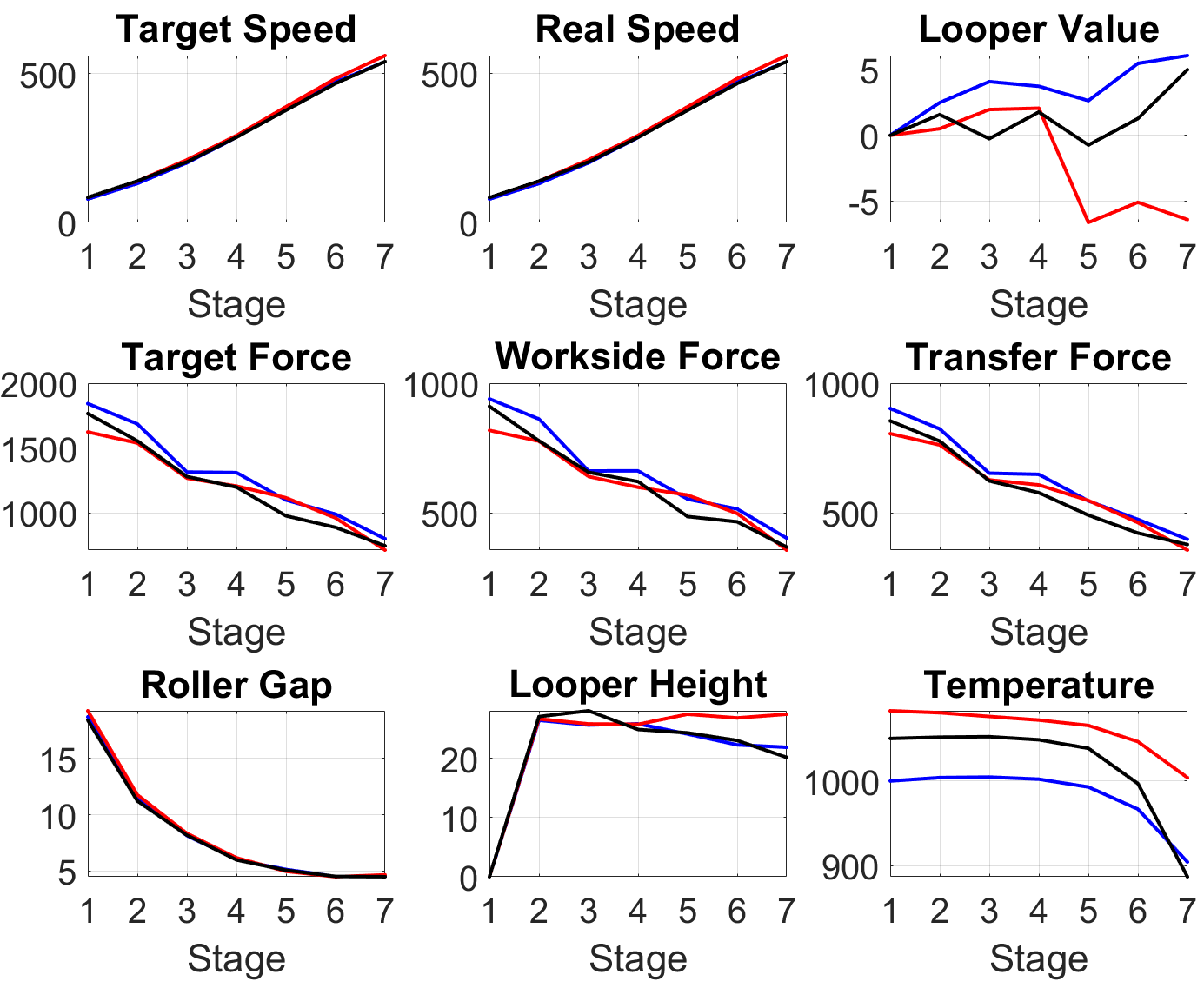} }}%
	\caption{Process variables for the hot strip mill (the three colors represent three products).}
	\label{fig:pv}
\end{figure}

Following the suggestion of the engineers who work in the field, we focus on 9 process variables: \textit{target speed of rollers}, \textit{the measured speed of rollers}, \textit{looper value}, \textit{target force on both side of the rollers}, \textit{the measured force on the work side of rollers}, \textit{measured force on the transfer side of rollers}, \textit{roller gap}, \textit{looper height}, and \textit{temperature}. At each stage, each of the process variable is a profile with $1,500$ data points observed during the steel strip rolling process (see Figure \ref{fig:pv}(a)). We take the average of each profile and set it as the value of the process variable at that stage. By doing so, each process variable is reduced to be one value at each stage. As a result, we construct a process variable matrix $\mathbf{X}_i$ with 9 rows and 7 columns for the $i$th product.

 Similar to the simulation study in Section \ref{sec:simulation}, we compare the performance of our proposed method to three benchmarking models: Row-Column Selection (benchmark I), Column-Row Selection (benchmark II), and Structured LASSO (benchmark III, \cite{Zhao2014}). To get a stable selection result, we randomly select $400$ samples from the dataset (which has $490$ samples in total) and construct a sub dataset. We then apply our method and the three benchmarks to the sub dataset to identify crucial rows and columns. We repeat this procedure $100$ times and then compute the selection percentage for each row and column of the process variable matrix. Any rows (or columns) whose selection rates are higher than 0.5 are considered as important rows (or columns). The selection results for process variables and stages are reported in Tables \ref{tb:a4} and \ref{tb:a5}, respectively.

\renewcommand{\arraystretch}{1}
\begin{table}[htb!]
	\begin{center}
		\footnotesize
		\begin{tabular}{c|cccc}
			\toprule
			Process variables & Proposed method & Benchmark I & Benchmark II & Benchmark III \\
			\midrule
			Target speed  & 2 & 100 & 100 & 3  \\ 
			\midrule 
			Measured speed  & 2 & 100 & 100 & 3 \\
			\midrule
			Looper value  & 100 & 100 & 100 & 100  \\
			\midrule
			Both side target force  & 0 & 0 & 0 & 0  \\
			\midrule
			Work side force  & 0 & 0 & 0 & 0   \\
			\midrule
			Transfer forse  & 0 & 0 & 0 & 0  \\
			\midrule
			Roller gap  & 100 & 100 & 100 & 100  \\
			\midrule
			Looper height  & 100 & 100 & 100 & 100  \\
			\midrule
			Temperature  & 0 & 0 & 0 & 0   \\
			\bottomrule
		\end{tabular}
	\end{center}
	\vspace{-2ex}
	\caption{Selection rates (\%) of the process variables.}
	\label{tb:a4}
\end{table}

\renewcommand{\arraystretch}{1}
\begin{table}[htb!]
	\begin{center}
		\footnotesize
		\begin{tabular}{c|c|c|c|c|c|c|c}
			\toprule
			Method & Stage 1 & Stage 2 & Stage 3 & Stage 4 & Stage 5 & Stage 6 & Stage 7  \\
			\midrule
			Proposed method  & 1 & 1 & 100 & 100 & 38 & 88 & 3 \\  
			Benchmark I  & 100 & 100 & 100 & 100 & 100 & 100 & 100 \\
			Benchmark II  & 100 & 100 & 100 & 100 & 100 & 100 & 100 \\
			Benchmark III  & 2 & 2 & 100 & 100 & 37 & 91 & 5 \\
			\bottomrule
		\end{tabular}
	\end{center}
	\vspace{-2ex}
	\caption{Selection rates (\%) of the stages.}
	\label{tb:a5}
\end{table}

Tables \ref{tb:a4} and \ref{tb:a5} indicate that benchmarks I and II have identified the same crucial process variables and stages. They have selected 5 crucial process variables (i.e., \textit{target speed}, \textit{measured speed}, \textit{looper value}, \textit{looper height}, and \textit{roller gap}) and all 7 stages. Our proposed method and the structured LASSO (benchmark III) have identified the same crucial process variables (\textit{looper value}, \textit{looper height}, and \textit{roller gap}) and stages (\textit{Stages 3, 4, 6}). This implies that benchmarks I and II are not efficient in identifying crucial process variables and stages since they select 5 process variables and all 7 stages, whereas our method and benchmark III have identified a subset of process variables and stages. Moreover, our method and benchmark III selecting the same crucial process variables and stages suggests that a penalized matrix regression model with a rank-one coefficient matrix is suitable to model the relationship between product quality defects and process variable matrix in this case study. We have discussed the selection results with some engineers in the steel company who provides us with the data and acquired reasonable explanations that why these process variables and stages are identified as important ones for product quality defects. To be specific, process variables \textit{looper value} and \textit{looper height} are used to control the tension of the steel strip between two stages. Process variable \textit{roller gap} is used to control the thickness of the steel strip, which also significantly affects the real-time value of \textit{looper value}. In other words, these three process variables are coupled and thus pose significant challenges for the closed-loop control system to adjust their values timely and correctly in reality. In addition, stages 1-3 of the hot roll mill respectively have a speed reducer connected to the rollers to reduce the speed from the driven motors, whereas stages 4-7 do not have any reducer. Therefore, the moving speed of the steel strip in stages 4-7 are much higher and thus it is more challenging for the real-time feedback control of the three identified process variables in stages 4-7. In other words, it is reasonable to select stages 4 and 6 as crucial stages, whereas the reason that stage 3 is also identified as a crucial stage needs further investigation.

\section{Conclusions}\label{sec:conclusion}

The advancements in sensing technology and data acquisition systems have facilitated us to collect a massive amount of control and sensor data during the operation of multistage processes. These data usually contain rich information of the processes and can be utilized for the root-cause diagnostics of product quality defects. The diagnostics of product quality defects in a multistage process often requires a simultaneous identification of both crucial stages and process variables. However, most of the existing variable selection methods such as LASSO, group LASSO, and their variants focus on one-dimensional variable selection. Although the crucial stages and process variables can still be identified by sequentially applying group LASSO to select the crucial stages first and then the crucial process variables, the selection accuracy is compromised since non-crucial process variables negatively affect the selection accuracy of important stages and vice versa. 

To address the challenge, this paper proposed a 2D variable selection methodology. The method regresses the product quality index against a matrix, whose rows represent process variables and columns are stages, using a generalized linear model. To simultaneously identify the crucial process variables and stages (i.e., rows and columns of the matrix respectively), we decompose the unknown regression coefficient matrix as a product of two factor matrices and penalize the rows of first factor matrix and the columns of second matrix using adaptive group LASSO. This yields an estimated coefficient matrix with both sparse rows and columns. The process variables (or stages) corresponding to the nonzero rows (or columns) of the coefficient matrix are considered as crucial process variables (or stages) that are responsible for the product quality defects. To estimate the sparse coefficient matrix, we have developed a block coordinate proximal descent optimization algorithm. The algorithm iteratively optimizes one factor matrix while keeping the other one fixed until convergence. We have proved that the proposed block coordinate proximal descent algorithm always converges to a critical point from any initialization point. In addition, we have also proved that each of the sub-optimization problems has a closed-form solution if the quality index follows a distribution whose (negative) log-likelihood function has a Lipschitz continuous gradient.

We have used numerical studies to validate the effectiveness of the proposed 2D variable selection method. The results from a simulation study indicate that our proposed method always achieves higher selection accuracy and precision than the designed benchmarks for data with various NSRs, sample sizes, and correlation structures. We believe this is because our method simultaneously selects the crucial rows and columns of a matrix while the benchmarks either sequentially select them or preserve an assumption that the rank of the regression coefficient matrix is one. A quality defect diagnostics dataset from the real-world steel industry has also been used to evaluate the performance of the proposed method. We identified \textit{looper value}, \textit{looper height}, and \textit{roller gap} as the crucial process variables and stage 3, 4 and 6 as the crucial stages. We have discussed the selection results with the engineers that work in the field and acquired reasonable explanations for the selection results.  

\newpage
\appendix
\section*{Appendices}
\addcontentsline{toc}{section}{Appendices}
\renewcommand{\thesubsection}{\Alph{subsection}}

\subsection{Proof for Proposition 1}

$\boldsymbol a^\top\mathbf{X}_i\boldsymbol b=(\boldsymbol b\otimes\boldsymbol a)^\top\text{vec}(\mathbf{X}_i)=\big\langle\boldsymbol b\otimes\boldsymbol a, \text{vec}(\mathbf{X}_i)\big\rangle=\big\langle\text{vec}(\boldsymbol a\boldsymbol b^\top), \text{vec}(\mathbf{X}_i)\big\rangle =\big\langle\text{vec}(\mathbf{B}), \text{vec}(\mathbf{X}_i)\big\rangle \\=\big\langle\mathbf{B}, \mathbf{X}_i\big\rangle$, where $\otimes$ is the Kronecker product.  \quad\qedsymbol
	
\subsection{Proof for Theorem 1}

Equations \eqref{eq:bcgdb1} and \eqref{eq:bcgdb2} can be easily proved using the gradient descent method. Therefore, we only prove Equations \eqref{eq:bcgdu} and \eqref{eq:bcgdv} here. The regularization term $ \mathcal{R}(\mathbf{U},\mathbf{V}) $ in Equation \eqref{eq:reg} can be decomposed as $ \mathcal{R}(\mathbf{U},\mathbf{V}) = \mathcal{R}_1(\mathbf{U}) + \mathcal{R}_2(\mathbf{V}) $, where 

\begin{equation}\nonumber
\begin{aligned}
\mathcal{R}_1(\mathbf{U}) = \lambda\gamma \sum_{j=1}^s\frac{\Vert\mathbf{u}_j\Vert}{\Vert\hat{\mathbf{u}}_j\Vert},
\end{aligned}
\end{equation}
\begin{equation}\nonumber
\begin{aligned}
\mathcal{R}_2(\mathbf{V}) = \lambda\gamma \sum_{k=1}^t\frac{\Vert\mathbf{v}_k\Vert}{\Vert\hat{\mathbf{v}}_k\Vert}.
\end{aligned}
\end{equation}
Here, $ \lambda \ge 0 $ is the tuning parameter. $ \Vert\cdot\Vert $ is $ \ell_2 $ norm. $ \gamma = \sqrt{r} $. $\mathbf{u}_j\in\mathbb{R}^r$ is the $j$th row of matrix $ \mathbf{U} $ and $\mathbf{v}_k\in\mathbb{R}^r$ is the $k$th column of matrix $ \mathbf{V}$. $ \hat{\mathbf{u}}_j $ and $ \hat{\mathbf{v}}_k $ are respectively the maximum likelihood estimates of $ \mathbf{u}_j $ and $\mathbf{v}_k$ when $ \lambda = 0 $. 

To derive the closed-form solutions for Equations \eqref{eq:gu} and \eqref{eq:gv}, we first introduce Moreau decomposition. Let prox($ \cdot $) be the proximal operator and $ \mathcal{R}_1^{*} $ be the conjugate of $ \mathcal{R}_1 $. Moreau decomposition for the term $ \mathcal{R}_1 $ is as follows:
\begin{equation}\nonumber
\begin{aligned}
\mathbf{U} = \text{prox}_{\mathcal{R}_1}(\mathbf{U})+\text{prox}_{\mathcal{R}_1^{*}}(\mathbf{U}).
\end{aligned}
\end{equation}
where 
\begin{equation}\nonumber
\begin{aligned}
\mathbf{U} = (\mathbf{u_1}^\intercal, \mathbf{u_2}^\intercal, ..., \mathbf{u_s}^\intercal)^\top,
\end{aligned}
\end{equation}
\begin{equation}\nonumber
\begin{aligned}
\text{prox}_{\mathcal{R}_1}(\mathbf{U}) = (\text{prox}_{\mathcal{R}_{11} }(\mathbf{u_1}), \text{prox}_{\mathcal{R}_{12}}(\mathbf{u_2}), ..., \text{prox}_{\mathcal{R}_{1s}}(\mathbf{u_s}))^\top,
\end{aligned}
\end{equation}
\begin{equation}\nonumber
\begin{aligned}
\text{prox}_{\mathcal{R}_1^{*}}(\mathbf{U}) = (\text{prox}_{\mathcal{R}_{11}^{*}}(\mathbf{u_1}), \text{prox}_{\mathcal{R}_{12}^{*}}(\mathbf{u_2}), ..., \text{prox}_{\mathcal{R}_{1s}^{*}}(\mathbf{u_s}))^\top.
\end{aligned}
\end{equation}
\noindent
We then represent the proximal operator of the conjugate as a projection operator $ \mathcal{P}(\cdot) $:
\begin{equation}\nonumber
\begin{aligned}
\text{prox}_{\mathcal{R}_{1j}}(\mathbf{u}_j)=\mathbf{u}_j-\mathcal{P}_{\mathcal{B}_{*}(\tau_u\gamma)}(\mathbf{u}_j),
\end{aligned}
\end{equation}
where 
\begin{equation}\nonumber
\begin{aligned}
\mathcal{R}_{1j} = \lambda\gamma\frac{\Vert{\mathbf{u}}_j\Vert}{ \Vert\hat{\mathbf{u}}_j\Vert},\quad  \forall j=1,2,...,s \quad \text{\textit{and}}
\end{aligned}
\end{equation}
\begin{equation}\nonumber
\begin{aligned}
\mathcal{B}_{*}(\tau_u\gamma) = \left\{\mathbf{u}_j\in\mathbb{R}^r : \Vert\mathbf{u}_j\Vert\le\tau_u\gamma,\ \ \text{where} \ \  \tau_u=\frac{\lambda}{ L_u^k \Vert\hat{\mathbf{u}}_j\Vert},\  \gamma=\sqrt{r},\  \forall j=1,2,...,s \right\}
\end{aligned}
\end{equation}
is the ball of dual norm. Thus, the proximal operator of the conjugate of the term $ \mathcal{R}_1 $ becomes a projection onto the ball of the dual norm, which is expressed by:

\begin{equation}\nonumber
\begin{aligned}
\mathcal{P}_{\mathcal{B}_{*}(\tau_u\gamma)}(\mathbf{u}_j) = 
\begin{cases}
\tau_u\gamma\frac{\mathbf{u}_j}{\Vert\mathbf{u}_j\Vert}, & \mbox{if } \Vert\mathbf{u}_j\Vert > \tau_u\gamma \\
\mathbf{u}_j, & \mbox{if } \Vert\mathbf{u}_j\Vert \le \tau_u\gamma
\end{cases}
\end{aligned}
\end{equation}

\noindent Therefore, we can derive a closed-form solution for Equation \eqref{eq:gu}, which is the row-wise soft-threshoding operator $ (\mathcal{S}_{\tau_u})_j $ shown below:

\begin{equation}\nonumber
\begin{aligned}
(\mathcal{S}_{\tau_u}(\mathbf{u}))_j = \text{prox}_{\mathcal{R}_1j}(\mathbf{u}_j) =
\begin{cases}
\mathbf{u}_j-\tau_u\gamma\frac{\mathbf{u}_j}{\Vert\mathbf{u}_j\Vert}, & \mbox{if } \Vert\mathbf{u}_j\Vert > \tau_u\gamma \\
\mathbf{0}, & \mbox{if } \Vert\mathbf{u}_j\Vert \le \tau_u\gamma
\end{cases}
\end{aligned}
\end{equation}

\noindent The column-wise soft-threshoding operator $ (\mathcal{S}_{\tau_v})_k $ for Equation \eqref{eq:gv} can be derived similarly. \quad\qedsymbol

\newpage
\subsection{Derivations of the Lipschitz constants}
For a Bernoulli distribution, we have the following:

\begin{equation}\nonumber
\begin{aligned}
\mathbb{P}(y_i|\mathbf{X}_i;\mathbf{U}, \mathbf{V}, \beta) 
&= p_i^{y_i}(1-p_i)^{1-y_i} \quad \text{where} \quad p_i = e^{\beta+\langle\mathbf{X}_i\mathbf{V}^\top,\mathbf{U}\rangle}/(1+e^{\beta+\langle\mathbf{X}_i\mathbf{V}^\top,\mathbf{U}\rangle})\\ &= \exp{\Big\{y_i\big(\beta+\big\langle\mathbf{X}_i\mathbf{V}^\top,\mathbf{U}\big\rangle\big)- \log{\big(1+\exp\big[\beta+\big\langle\mathbf{X}_i\mathbf{V}^\top,\mathbf{U}\big\rangle\big]\big)\Big\}}}, \\ 
\mathcal{L}\big(\mathbf{U},\mathbf{V},\beta\big) &= -\sum_{i=1}^n \Big\{y_i\big(\beta+\big\langle\mathbf{X}_i\mathbf{V}^\top,\mathbf{U}\big\rangle\big) - \log{\big(1+\exp\big[\beta+\big\langle\mathbf{X}_i\mathbf{V}^\top,\mathbf{U}\big\rangle\big]\big)\Big\}}, \\
\nabla_{(\mathbf{U},\beta)}\mathcal{L}\big(\mathbf{U},\mathbf{V},\beta\big) &=
\nabla_{(\mathbf{U},\beta)}\left\{-\sum_{i=1}^n \Big(y_i\big(\beta+\big\langle\mathbf{X}_i\mathbf{V}^\top,\mathbf{U}\big\rangle\big) - \log{\big(1+\exp\big[\beta+\big\langle\mathbf{X}_i\mathbf{V}^\top,\mathbf{U}\big\rangle\big]\big)\Big)}\right\} \\
&= -\sum_{i=1}^n \Big(y_i-\big(1+\exp{\big[-\beta-\big\langle\mathbf{X}_i\mathbf{V}^\top,\mathbf{U}\big\rangle\big]}\big)^{-1}\Big)\big(1+\mathbf{X}_i\mathbf{V}^\top\big).
\end{aligned}
\end{equation}

\begin{flalign}\nonumber
&\begin{aligned}
&\;\; \Vert\nabla_{(\mathbf{U},\beta)}\mathcal{L}\big(\mathbf{U},\mathbf{V},\beta\big)-\nabla_{(\mathbf{U},\beta)}\mathcal{L}\big(\tilde{\mathbf{U}},\mathbf{V},\tilde{\beta}\big)\Vert_F \\
&= \Big\Vert  -\sum_{i=1}^n \left\{\big(1+\exp{\big[-\beta-\big\langle\mathbf{X}_i\mathbf{V}^\top,\mathbf{U}\big\rangle\big]}\big)^{-1}-\big(1+\exp{\big[-\tilde{\beta}-\big\langle\mathbf{X}_i\mathbf{V}^\top,\tilde{\mathbf{U}}\big\rangle\big]}\big)^{-1}\right\}\big(1+\mathbf{X}_i\mathbf{V}^\top\big)\Big\Vert_F \\
&\le \sum_{i=1}^n \left|\big(1+\exp{\big[-\beta-\big\langle\mathbf{X}_i\mathbf{V}^\top,\mathbf{U}\big\rangle\big]}\big)^{-1}-\big(1+\exp{\big[-\tilde{\beta}-\big\langle\mathbf{X}_i\mathbf{V}^\top,\tilde{\mathbf{U}}\big\rangle\big]}\big)^{-1}\right| \Vert 1+\mathbf{X}_i\mathbf{V}^\top\Vert_F \\
&\le \sum_{i=1}^n \left|-\beta-\big\langle\mathbf{X}_i\mathbf{V}^\top,\mathbf{U}\big\rangle+\tilde{\beta}+\big\langle\mathbf{X}_i\mathbf{V}^\top,\tilde{\mathbf{U}}\big\rangle\right|\Vert 1+\mathbf{X}_i\mathbf{V}^\top\Vert_F \\
&\le \sum_{i=1}^n
\left(\left|\big\langle\mathbf{X}_i\mathbf{V}^\top,\mathbf{U}-\tilde{\mathbf{U}}\big\rangle\right|+|\beta-\tilde{\beta}|\right)\Vert 1+\mathbf{X}_i\mathbf{V}^\top\Vert_F \\
&\le \sum_{i=1}^n
\left(\Vert\mathbf{X}_i\mathbf{V}^\top\Vert_F\Vert\mathbf{U}-\tilde{\mathbf{U}}\Vert_F+|\beta-\tilde{\beta}|\right)\Vert 1+\mathbf{X}_i\mathbf{V}^\top\Vert_F \\
&\le \sum_{i=1}^n
\left(\Vert\mathbf{U}-\tilde{\mathbf{U}}\Vert_F+|\beta-\tilde{\beta}|\right)\left(1+\Vert \mathbf{X}_i\mathbf{V}^\top\Vert_F\right)\Vert 1+\mathbf{X}_i\mathbf{V}^\top\Vert_F \\
&\le \sqrt{2}\ \sum_{i=1}^n
\left(1+\Vert \mathbf{X}_i\mathbf{V}^\top\Vert_F\right)\Vert 1+\mathbf{X}_i\mathbf{V}^\top\Vert_F
\Vert\big(\mathbf{U},\beta\big)-\big(\tilde{\mathbf{U}},\tilde{\beta}\big)\Vert_F.
\end{aligned} &&
\end{flalign}
where in the third inequality we used the fact that 
\begin{equation}\nonumber
\begin{aligned}
|(1+e^x)^{-1}-(1+e^y)^{-1}| \le |x-y|,
\end{aligned}
\end{equation}
and in the seventh inequality we used the Cauchy-Schwarz inequality:
\begin{equation}\nonumber
\begin{aligned}
\big\Vert\mathbf{U}-\tilde{\mathbf{U}}\big\Vert_F+\big|\beta-\tilde{\beta}\big| \le \sqrt{2}\big\Vert\big(\mathbf{U},\beta\big)-\big(\tilde{\mathbf{U}},\tilde{\beta}\big)\Vert_F.
\end{aligned}
\end{equation}
The constant $ L_v $ can be derived similarly. \quad\qedsymbol \\

\noindent For a binomial distribution, its Lipschitz constant is almost the same as the Bernoulli distribution except for an additional $ n_i $ term.

\begin{equation}\nonumber
\begin{aligned}
\mathbb{P}(y_i|\mathbf{X}_i;\mathbf{U}, \mathbf{V}, \beta) 
&= {n_i \choose y_i} p_i^{y_i}(1-p_i)^{n_i-y_i} \quad \text{where} \quad p_i = e^{\beta+\langle\mathbf{X}_i\mathbf{V}^\top,\mathbf{U}\rangle}/(1+e^{\beta+\langle\mathbf{X}_i\mathbf{V}^\top,\mathbf{U}\rangle})\\ &= {n_i \choose y_i}\exp{\Big\{y_i\big(\beta+\big\langle\mathbf{X}_i\mathbf{V}^\top,\mathbf{U}\big\rangle\big)- n_i \log{\big(1+\exp\big[\beta+\big\langle\mathbf{X}_i\mathbf{V}^\top,\mathbf{U}\big\rangle\big]\big)\Big\}}}.
\end{aligned}
\end{equation}

\begin{flalign}\nonumber
&\begin{aligned}
&\;\; \Vert\nabla_{(\mathbf{U},\beta)}\mathcal{L}\big(\mathbf{U},\mathbf{V},\beta\big)-\nabla_{(\mathbf{U},\beta)}\mathcal{L}\big(\tilde{\mathbf{U}},\mathbf{V},\tilde{\beta}\big)\Vert_F \\
&\qquad\quad\le \sqrt{2}\ \sum_{i=1}^n n_i
\left(1+\Vert \mathbf{X}_i\mathbf{V}^\top\Vert_F\right)\Vert 1+\mathbf{X}_i\mathbf{V}^\top\Vert_F
\Vert\big(\mathbf{U},\beta\big)-\big(\tilde{\mathbf{U}},\tilde{\beta}\big)\Vert_F. \quad\qedsymbol
\end{aligned} &&
\end{flalign}  \\
\noindent For a normal distribution, we have the following:
\begin{equation}\nonumber
\begin{aligned} 
\mathbb{P}(y_i|\mathbf{X}_i;\mathbf{U},\mathbf{V}, \beta) &= \frac{1}{\sqrt{2\pi}\sigma}\exp{\left\{-\frac{\big(y_i-(\beta+\langle\mathbf{X}_i\mathbf{V}^\top,\mathbf{U}\big\rangle\big)^2}{2\sigma^2}\right\}},
\\
\mathcal{L}\big(\mathbf{U},\mathbf{V},\beta\big) &= \frac{n}{2}\log(2\pi\sigma^2) -\sum_{i=1}^n \left\{-\frac{y_i^2}{2\sigma^2}+\frac{1}{\sigma^2}y_i(\beta+\langle\mathbf{X}_i\mathbf{V}^\top,\mathbf{U}\big\rangle\big)-\frac{1}{2\sigma^2}\big(\beta+\langle\mathbf{X}_i\mathbf{V}^\top,\mathbf{U}\big\rangle\big)^2\right\},
\\
\nabla_{(\mathbf{U},\beta)}\mathcal{L}\big(\mathbf{U},\mathbf{V},\beta\big) &=
-\frac{1}{\sigma^2}\sum_{i=1}^n\Big\{y_i\big(1+\mathbf{X}_i\mathbf{V}^\top  \big) - \big(\beta+\langle\mathbf{X}_i\mathbf{V}^\top,\mathbf{U}\big\rangle\big)\big(1+\mathbf{X}_i\mathbf{V}^\top \big)\Big\} \\
&= -\frac{1}{\sigma^2}\sum_{i=1}^n \big(y_i-\beta-\big\langle\mathbf{X}_i\mathbf{V}^\top,\mathbf{U}\big\rangle\big)\big(1+\mathbf{X}_i\mathbf{V}^\top\big).
\end{aligned}
\end{equation}

\begin{flalign}\nonumber
&\begin{aligned}
&\;\; \Vert\nabla_{(\mathbf{U},\beta)}\mathcal{L}\big(\mathbf{U},\mathbf{V},\beta\big)-\nabla_{(\mathbf{U},\beta)}\mathcal{L}\big(\tilde{\mathbf{U}},\mathbf{V},\tilde{\beta}\big)\Vert_F \\
&\qquad\quad= \Big\Vert  -\frac{1}{\sigma^2}\sum_{i=1}^n \big(y_i-\beta-\big\langle\mathbf{X}_i\mathbf{V}^\top,\mathbf{U}\big\rangle-y_i+\tilde{\beta}+\big\langle\mathbf{X}_i\mathbf{V}^\top,\tilde{\mathbf{U}}\big\rangle\big)\big(1+\mathbf{X}_i\mathbf{V}^\top\big)\Big\Vert_F \\
&\qquad\quad\le \frac{1}{\sigma^2}\sum_{i=1}^n \left|-\beta-\big\langle\mathbf{X}_i\mathbf{V}^\top,\mathbf{U}\big\rangle+\tilde{\beta}+\big\langle\mathbf{X}_i\mathbf{V}^\top,\tilde{\mathbf{U}}\big\rangle\right|\Vert 1+\mathbf{X}_i\mathbf{V}^\top\Vert_F \\
&\qquad\quad\le \frac{1}{\sigma^2}\sum_{i=1}^n
\left(\left|\big\langle\mathbf{X}_i\mathbf{V}^\top,\mathbf{U}-\tilde{\mathbf{U}}\big\rangle\right|+|\beta-\tilde{\beta}|\right)\Vert 1+\mathbf{X}_i\mathbf{V}^\top\Vert_F \\
&\qquad\quad\le \frac{1}{\sigma^2}\sum_{i=1}^n
\left(\Vert\mathbf{X}_i\mathbf{V}^\top\Vert_F\Vert\mathbf{U}-\tilde{\mathbf{U}}\Vert_F+|\beta-\tilde{\beta}|\right)\Vert 1+\mathbf{X}_i\mathbf{V}^\top\Vert_F \\
&\qquad\quad\le \frac{1}{\sigma^2}\sum_{i=1}^n
\left(\Vert\mathbf{U}-\tilde{\mathbf{U}}\Vert_F+|\beta-\tilde{\beta}|\right)\left(1+\Vert \mathbf{X}_i\mathbf{V}^\top\Vert_F\right)\Vert 1+\mathbf{X}_i\mathbf{V}^\top\Vert_F \\
&\qquad\quad\le \frac{\sqrt{2}}{\sigma^2}\sum_{i=1}^n
\left(1+\Vert \mathbf{X}_i\mathbf{V}^\top\Vert_F\right)\Vert 1+\mathbf{X}_i\mathbf{V}^\top\Vert_F
\Vert\big(\mathbf{U},\beta\big)-\big(\tilde{\mathbf{U}},\tilde{\beta}\big)\Vert_F.
\end{aligned} &&
\end{flalign}
where in the second inequality we used the same fact as the one used in the Bernoulli distribution case.\quad\qedsymbol

\subsection{Proof for Theorem 2}
Theorem 2 can be proved following procedures similar to the ones in \cite{Shi2014} and \cite{Xu2013}, which establish the global convergence of the cyclic block coordinate proximal method by assuming that the Kurdyka-Lojasiewicz (KL) inequality holds. We will first show that the proposed BCPD algorithm possesses the subsequence convergence property, which implies that a limit point of the sequence is a stationary point (critical point) of problem \eqref{eq:opt}. In addition, we will show that the objective function in \eqref{eq:opt} is sub-analytic and thus satisfies the KL inequality. Given the inequality holds, we will prove that if we have a sequence generated by the proposed BCPD algorithm close to a limit point, then the sequence converges to this limit point and thus a critical point.

We start with the following assumptions.

\begin{assumptionC}
Let $ F $ be continuous in dom($ F$). We assume the infimum of $ F(\mathbf{U},\mathbf{V},\beta) $ exists and the problem \eqref{eq:opt} has at least one stationary point. In addition, we assume the sequence $ \{(\mathbf{U}^k,\mathbf{V}^k,\beta^k)\} $ is bounded. Each block of $( \mathbf{U}^k,\mathbf{V}^k,\beta^k)$ is updated by Equations \eqref{eq:gu}-\eqref{eq:gb2}. Furthermore, there exist constants $ 0 < \ell_u \le L_u < \infty $ and $ 0 < \ell_v \le L_v < \infty $ such that  $ L_u^{k}$ and $L_v^{k}$ satisfy $ \ell_u \le L_u^{k} < L_u$ and $ \ell_v \le L_v^{k} < L_v$, respectively.
\end{assumptionC}
Note that if $ \nabla_{\mathbf{U}}\mathcal{L}(\mathbf{U},\mathbf{V},\beta) $ and $ \nabla_{\mathbf{V}}\mathcal{L}(\mathbf{U},\mathbf{V},\beta) $ are Lipschitz continous, it is obvious that $ L_u^{k}$ and $ L_v^{k}$ are bounded according to \eqref{eq:luconst} and \eqref{eq:lvconst} because $ \{(\mathbf{U}^k,\mathbf{V}^k,\beta^k)\} $ is bounded. In general, $ L_u^{k}$ (and $ L_v^{k}$) is not necessarily the Lipschitz constants of $ \nabla_{\mathbf{U}}\mathcal{L}(\mathbf{U},\mathbf{V},\beta) $ (and $ \nabla_{\mathbf{V}}\mathcal{L}(\mathbf{U},\mathbf{V},\beta) $). However, both of them are required to be uniformly lower bounded from zero and upper bounded.  See more details in \cite{Shi2014, Xu2013}. 

Given Assumption C1, we have the following lemma.

\begin{lemmaC}[Subsequence Convergence]
	Under Assumption C1, let $ \{(\mathbf{U}^k,\mathbf{V}^k,\beta^k)\} $ be the sequence generated by the proposed BCPD algorithm. Then, 
	any limit point $ (\bar{\mathbf{U}},\bar{\mathbf{V}},\bar{\beta}) $ of $ \{(\mathbf{U}^k,\mathbf{V}^k,\beta^k)\} $ is a stationary point of \eqref{eq:opt}.
\end{lemmaC}

\noindent \textbf{Proof}. Let $ F(\mathbf{U}^k,\mathbf{V}^k,\beta^k) = \mathcal{L}(\mathbf{U}^k,\mathbf{V}^k,\beta^k) + \mathcal{R}(\mathbf{U}^k,\mathbf{V}^k) $ be the value of the objective function in \eqref{eq:opt} at $ (\mathbf{U}^k,\mathbf{V}^k,\beta^k) $. According to Lemma 2.3 in \cite{Beck2009} and Lemma 2.1 in \cite{Xu2013}, if $ f_1(\mathbf{x}^{*}) \le f_1(\mathbf{y}) + \langle \nabla f_1(\mathbf{y}), \mathbf{x}^{*} - \mathbf{y} \rangle + \frac{L}{2} \Vert \mathbf{x}^{*} - \mathbf{y} \Vert^2 $, then $ f(\mathbf{x}) - f(\mathbf{x}^{*}) \ge \frac{L}{2} \Vert \mathbf{x}^{*} - \mathbf{y} \Vert^2 + L \langle \mathbf{y} - \mathbf{x}, \mathbf{x}^{*} - \mathbf{y} \rangle $ for any $ \mathbf{x} \in \mathcal{X} $ under the settings that $ f_1(\mathbf{x}) $ and $ f_2(\mathbf{x}) $ are two convex functions defined on the convex set $ \mathcal{X} $, $ f_1(\mathbf{x}) $ is differentiable, $ f(\mathbf{x}) = f_1(\mathbf{x}) + f_2(\mathbf{x}) $, and $ \mathbf{x}^{*} = \argmin_{\mathbf{x} \in \mathcal{X}} \langle \nabla f_1(\mathbf{x}), \mathbf{x} - \mathbf{y} \rangle + \frac{L}{2} \Vert \mathbf{x} - \mathbf{y} \Vert^2 + f_2(\mathbf{x}).$ Here, \\

\noindent i) Set $ \mathbf{x} = \mathbf{y} =  (\mathbf{U}^{k-1},\mathbf{V}^{k-1},\beta^{k-1}) $. Then, $ \mathbf{x}^{*} = (\mathbf{U}^k,\mathbf{V}^{k-1},\hat{\beta}^k) $. We have

\begin{flalign}\nonumber
&\begin{aligned}
\;
\qquad F(\mathbf{U}^{k-1},\mathbf{V}^{k-1},\beta^{k-1}) - F(\mathbf{U}^k,\mathbf{V}^{k-1},\hat{\beta}^k)
& \ge \frac{L_u^{k}}{2} \Vert (\mathbf{U}^k,\mathbf{V}^{k-1},\hat{\beta}^k) - (\mathbf{U}^{k-1},\mathbf{V}^{k-1},\beta^{k-1}) \Vert^2_F \\
& = \frac{L_u^{k}}{2} \Vert (\mathbf{U}^k-\mathbf{U}^{k-1}, \mathbf{V}^{k-1}-\mathbf{V}^{k-1}, \hat{\beta}^k-\beta^{k-1}) \Vert^2_F \\ 
& =
\frac{L_u^{k}}{2} \big(\Vert\mathbf{U}^{k-1}-{\mathbf{U}}^k\Vert^2_F + 
\lvert \beta^{k-1} - \hat{\beta}^k\rvert^2 \big),
\end{aligned} &&
\end{flalign}

\noindent ii) Set $ \mathbf{x} = \mathbf{y} =  (\mathbf{U}^k,\mathbf{V}^{k-1},\hat{\beta}^k) $. Then, $ \mathbf{x}^{*} = (\mathbf{U}^k,\mathbf{V}^{k},{\beta}^k) $. We have

\begin{flalign}\nonumber
&\begin{aligned}
\;
\qquad F(\mathbf{U}^k,\mathbf{V}^{k-1},\hat{\beta}^k) - F(\mathbf{U}^k,\mathbf{V}^{k},{\beta}^k)
& \ge \frac{L_v^{k}}{2} \Vert (\mathbf{U}^k,\mathbf{V}^{k},{\beta}^k)  -  (\mathbf{U}^k,\mathbf{V}^{k-1},\hat{\beta}^k) \Vert^2_F\\
& = \frac{L_v^{k}}{2} \Vert (\mathbf{U}^k-\mathbf{U}^k, \mathbf{V}^{k}-\mathbf{V}^{k-1}, {\beta}^k-\hat{\beta}^k) \Vert^2_F \\ 
& =
\frac{L_v^{k}}{2} \big(\Vert\mathbf{V}^{k-1}-{\mathbf{V}}^k\Vert^2_F + 
\lvert \hat{\beta}^{k} - \beta^k\rvert^2 \big).
\end{aligned} &&
\end{flalign}

where $ \Vert (\mathbf{U},\mathbf{V},\beta)\Vert_F = \sqrt{\Vert\mathbf{U}\Vert_F^2+\Vert\mathbf{V}\Vert_F^2+\beta^2} $. \\

\noindent Therefore, we sum up the inequalities in i) and ii) as follows:
\begin{flalign}\label{eq:Finequal}
&\begin{aligned}
\; 
\qquad F(\mathbf{U}^{k-1},\mathbf{V}^{k-1},\beta^{k-1}) - F(\mathbf{U}^k,\mathbf{V}^{k},\beta^k) & \ge 
\frac{L_u^{k}}{2} \big(\Vert\mathbf{U}^{k-1}-{\mathbf{U}}^k\Vert^2_F + 
\lvert \beta^{k-1} - \hat{\beta}^k\rvert^2 \big) \\ & \quad \; \; \ +
\frac{L_v^{k}}{2} \big(\Vert\mathbf{V}^{k-1}-{\mathbf{V}}^k\Vert^2_F + 
\lvert \hat{\beta}^{k} - \beta^k\rvert^2 \big).
\end{aligned} &&
\end{flalign}

\noindent Summing up the above inequality over $ k $ from 1 to $ K $ yields

\begin{flalign}\nonumber
&\begin{aligned}
F(\mathbf{U}^0,\mathbf{V}^0,\beta^0) - F(\mathbf{U}^K,\mathbf{V}^K,\beta^K) & \ge \sum_{k=1}^K \Big\{ \frac{L_u^{k}}{2} \big(\Vert\mathbf{U}^{k-1}-{\mathbf{U}}^k\Vert^2_F + 
\lvert \beta^{k-1} - \hat{\beta}^k\rvert^2 \big) \\ & \quad \; \; \ +
\frac{L_v^{k}}{2} \big(\Vert\mathbf{V}^{k-1}-{\mathbf{V}}^k\Vert^2_F + 
\lvert \hat{\beta}^{k} - \beta^k\rvert^2 \big) \Big\} \\ 
& \ge \sum_{k=1}^K \Big\{ \frac{\ell_u}{2} \big(\Vert\mathbf{U}^{k-1}-{\mathbf{U}}^k\Vert^2_F + 
\lvert \beta^{k-1} - \hat{\beta}^k\rvert^2 \big) \\ & \quad \; \; \ +
\frac{\ell_v}{2} \big(\Vert\mathbf{V}^{k-1}-{\mathbf{V}}^k\Vert^2_F + 
\lvert \hat{\beta}^{k} - \beta^k\rvert^2 \big) \Big\} \\ & \ge \sum_{k=1}^K \frac{\min(\ell_u,\ell_v)}{2} \big(\Vert\mathbf{U}^{k-1}-{\mathbf{U}}^k\Vert^2_F + 
\lvert \beta^{k-1} - \hat{\beta}^k\rvert^2 \\ & \quad \; \; \ +
\Vert\mathbf{V}^{k-1}-{\mathbf{V}}^k\Vert^2_F + 
\lvert \hat{\beta}^{k} - \beta^k\rvert^2 \big).
\end{aligned} &&
\end{flalign}

\noindent Since $ F $ is lower bounded, taking $ K \rightarrow \infty $ yields the square summable property as follows:
\begin{equation}\nonumber
\begin{aligned}
\sum_{k=1}^{\infty} \big(\Vert\mathbf{U}^{k-1}-{\mathbf{U}}^k\Vert^2_F + 
\Vert\mathbf{V}^{k-1}-{\mathbf{V}}^k\Vert^2_F + 
\lvert \beta^{k-1} - \hat{\beta}^k\rvert^2 +
\lvert \hat{\beta}^{k} - \beta^k\rvert^2 \big) < \infty.
\end{aligned}
\end{equation}

\noindent This inequality implies $ \Vert\mathbf{U}^{k-1}-{\mathbf{U}}^k\Vert^2_F + 
\Vert\mathbf{V}^{k-1}-{\mathbf{V}}^k\Vert^2_F + 
\lvert \beta^{k-1} - \hat{\beta}^k\rvert^2 +
\lvert \hat{\beta}^{k} - \beta^k\rvert^2 \rightarrow 0 $, which also means that $ \mathbf{U}^{k-1}-{\mathbf{U}}^k \rightarrow \bf{0} $, $ \mathbf{V}^{k-1}-{\mathbf{V}}^k \rightarrow \bf{0} $, $ \beta^{k-1} - \hat{\beta}^k \rightarrow 0 $, and $ \hat{\beta}^{k} - \beta^k \rightarrow 0 $. The last two results can be combined as $ \beta^{k-1} - \beta^k \rightarrow 0 $. Hence, 

\begin{equation} \label{eq:subseq}
\begin{aligned} (\mathbf{U}^{k},\mathbf{V}^{k},\beta^{k})-(\mathbf{U}^{k-1},\mathbf{V}^{k-1},\beta^{k-1}) \rightarrow \bf{0}. 
\end{aligned}
\end{equation}  

Now, let $ (\bar{\mathbf{U}},\bar{\mathbf{V}},\bar{\beta}) $ be a limit point of $ \{(\mathbf{U}^k,\mathbf{V}^k,\beta^k)\} $. Then, there exists a subsequence $ \{(\mathbf{U}^{k_j},\mathbf{V}^{k_j},\beta^{k_j})\} $ that converges to $ (\bar{\mathbf{U}},\bar{\mathbf{V}},\bar{\beta}) $. From \eqref{eq:subseq}, we know that
$ \{(\mathbf{U}^{k_j-1},\mathbf{V}^{k_j-1},\beta^{k_j-1})\} $ also converges to $ (\bar{\mathbf{U}},\bar{\mathbf{V}},\bar{\beta}) $.
 Since $ \{L_u^k\} $ and $ \{L_v^k\} $ are bounded, $ \{L_u^{k_j}\} $ and $ \{L_v^{k_j}\} $ converges to $ \{\bar{L}_u\} $ and $ \{\bar{L}_v\} $, respectively. Therefore, it is obvious that $ \beta^{k_j} \rightarrow \bar{\beta} $. From \eqref{eq:gu} and \eqref{eq:gv}, we have 
 
 \begin{equation}\nonumber
 \begin{aligned}
 \mathbf{U}^{k_j} = \argmin_{\mathbf{U}}\big\langle\nabla_\mathbf{U}\mathcal{L}\big(\mathbf{U}^{k_j-1},\mathbf{V}^{k_j-1},\beta^{k_j-1}\big), \mathbf{U}-\mathbf{U}^{k_j}\big\rangle +\frac{L_u^{k_j}}{2}\Vert
 \mathbf{U}-\mathbf{U}^{k_j-1}\Vert^2_F + \mathcal{R}(\mathbf{U},\mathbf{V}^{k_j-1}),
 \end{aligned}
 \end{equation}
 \begin{equation}\nonumber
 \begin{aligned}
 \mathbf{V}^{k_j} = \argmin_{\mathbf{V}}\big\langle\nabla_\mathbf{V}\mathcal{L}\big(\mathbf{U}^{k_j},\mathbf{V}^{k_j-1},\hat{\beta}^{k_j}\big), \mathbf{V}-\mathbf{V}^{k_j-1}\big\rangle +\frac{L_v^{k_j}}{2}\Vert
 \mathbf{V}-\mathbf{V}^{k_j-1}\Vert^2_F + \mathcal{R}(\mathbf{U}^{k_j},\mathbf{V}).
 \end{aligned}
 \end{equation} 
  
\noindent Let $ j  \rightarrow \infty $, we have 

\begin{equation}\nonumber
\begin{aligned}
\bar{\mathbf{U}} = \argmin_{\mathbf{U}}\big\langle\nabla_\mathbf{U}\mathcal{L}\big(\bar{\mathbf{U}},\bar{\mathbf{V}},\bar{\beta}\big), \mathbf{U}-\bar{\mathbf{U}}\big\rangle +\frac{\bar{L}_u}{2}\Vert
\mathbf{U}-\bar{\mathbf{U}}\Vert^2_F + \mathcal{R}(\mathbf{U},\bar{\mathbf{V}}),
\end{aligned}
\end{equation}
\begin{equation}\nonumber
\begin{aligned}
\bar{\mathbf{V}} = \argmin_{\mathbf{V}}\big\langle\nabla_\mathbf{V}\mathcal{L}\big(\bar{\mathbf{U}},\bar{\mathbf{V}},\bar{\beta}\big), \mathbf{V}-\bar{\mathbf{V}}\big\rangle +\frac{\bar{L}_v}{2}\Vert
\mathbf{V}-\bar{\mathbf{V}}\Vert^2_F + \mathcal{R}(\bar{\mathbf{U}},\mathbf{V}).
\end{aligned}
\end{equation} 
\noindent These equations imply $ \bf{0} \in \nabla_\mathbf{U}\mathcal{L}\big(\bar{\mathbf{U}},\bar{\mathbf{V}},\bar{\beta}\big) + \partial\mathcal{R}(\bar{\mathbf{U}},\bar{\mathbf{V}}) $ and $ \bf{0} \in \nabla_\mathbf{V}\mathcal{L}\big(\bar{\mathbf{U}},\bar{\mathbf{V}},\bar{\beta}\big) + \partial\mathcal{R}(\bar{\mathbf{U}},\bar{\mathbf{V}}) $. From \eqref{eq:gb1} and \eqref{eq:gb2}, we can also show that $ \nabla_\beta\mathcal{L}\big(\bar{\mathbf{U}},\bar{\mathbf{V}},\bar{\beta}\big) = 0 $ in a similar manner. Therefore, $ (\bar{\mathbf{U}},\bar{\mathbf{V}},\bar{\beta}) $ is a stationary point and thus a critical point of \eqref{eq:opt}. \quad\qedsymbol \\ 

To show the global convergence property of the proposed BCPD algorithm, we use the Kurdyka-Lojasiewicz inequality \cite{Shi2014, Xu2013}.

\begin{definitionC}[Kurdyka-Lojasiewicz Inequality]
A function $F(\mathbf{U},\mathbf{V},\beta)$ satisfies the Kurdyka-Lojasiewicz (KL) inequality at point $ F(\bar{\mathbf{U}},\bar{\mathbf{V}},\bar{\beta}) \in $ dom($\partial F$)  if there exists $ \theta \in [0,1) $ such that \\
\begin{equation} \label{eq:kl}
\frac{\lvert F(\mathbf{U},\mathbf{V},\beta) - F(\bar{\mathbf{U}},\bar{\mathbf{V}},\bar{\beta})\rvert^\theta} {\text{dist}(\mathbf{0},\partial F(\mathbf{U},\mathbf{V},\beta))}
\end{equation} \\
is bounded around $ (\bar{\mathbf{U}},\bar{\mathbf{V}},\bar{\beta}) $.
Namely, in a certain neiborhood $ \mathcal{N} $ of $ (\bar{\mathbf{U}},\bar{\mathbf{V}},\bar{\beta}) $, there exists $ \phi(s) = cs^{1-\theta} $ for some $ c > 0 $ and $ \theta \in [0,1) $ such that the KL inequality holds
\begin{equation}
\phi'(\lvert F(\mathbf{U},\mathbf{V},\beta) - F(\bar{\mathbf{U}},\bar{\mathbf{V}},\bar{\beta}) \rvert)\text{dist}(\mathbf{0},\partial F(\mathbf{U},\mathbf{V},\beta)) \ge 1, \\
\end{equation}
for any $(\mathbf{U},\mathbf{V},\beta) \in \mathcal{N} \cap \text{dom}(\partial F)\ and\ F(\mathbf{U},\mathbf{V},\beta) \neq F(\bar{\mathbf{U}},\bar{\mathbf{V}},\bar{\beta}) $,
where dom($\partial F$) $ \equiv \{(\mathbf{U},\mathbf{V},\beta): \partial F(\mathbf{U},\mathbf{V},\beta) \neq \emptyset \}$ and dist($\mathbf{0},\partial F(\mathbf{U},\mathbf{V},\beta)) \equiv \min\{\Vert \mathbf{Y} \Vert_F : \mathbf{Y} \in \partial F(\mathbf{U},\mathbf{V},\beta)\}$. 
\end{definitionC}

Note that \eqref{eq:kl} with $ \theta \in [1/2,1) $ is bounded around any critical point $ (\bar{\mathbf{U}},\bar{\mathbf{V}},\bar{\beta}) $ for real analytic functions \cite{Xu2013}. \eqref{eq:kl} with $ \theta \in [0,1) $ is also bounded around any critical point $ (\bar{\mathbf{U}},\bar{\mathbf{V}},\bar{\beta}) $ for  nonsmooth sub-analytic functions \cite{Bolte2007}. Therefore, both a real analytic function and a nonsmooth sub-analytic function satisfy the KL inequality. It is also known that i) both real analytic and semi-algebraic functions are sub-analytic; ii) the finite sum of real analytic functions is real analytic; iii) the finite sum of semi-algebraic functions are semi-algebraic; iv) the sum of real analytic functions and semi-algebraic functions is sub-analytic \cite{Bochnak1998, Xu2013}. For our optimization problem in \eqref{eq:opt}, $ \mathcal{L}(\mathbf{U},\mathbf{V},\beta) $ is a real analytic function. $ \mathcal{R}(\mathbf{U},\mathbf{V}) $ is a semi-algebraic function since the Euclidean norm $ \Vert \cdot \Vert $ is semi-algebraic \cite{Bochnak1998, Xu2013}. 
Therefore, the sum $ F(\mathbf{U},\mathbf{V},\beta) = \mathcal{L}(\mathbf{U},\mathbf{V},\beta) + \mathcal{R}(\mathbf{U},\mathbf{V}) $ is sub-analytic and thus satisfies the KL inequality. 

\begin{theoremC}[Global Convergence]
	Given Assumption C1 and the fact that $ F $ satisfies the KL inequality at a limit point $ (\bar{\mathbf{U}},\bar{\mathbf{V}},\bar{\beta}) $ of $ \{(\mathbf{U}^k,\mathbf{V}^k,\beta^k)\} $, then the sequence $ \{(\mathbf{U}^k,\mathbf{V}^k,\beta^k)\} $ converges to $ (\bar{\mathbf{U}},\bar{\mathbf{V}},\bar{\beta}) $, which is a critical point of \eqref{eq:opt}.
\end{theoremC}
\noindent \textbf{Proof.} 
Let $ (\bar{\mathbf{U}},\bar{\mathbf{V}},\bar{\beta}) $ be a limit point of $ \{(\mathbf{U}^k,\mathbf{V}^k,\beta^k)\} $. From the above statement, $ F $ satisfies KL inequality within a ball $ \mathcal{B}_\rho(\bar{\mathbf{U}},\bar{\mathbf{V}},\bar{\beta}) \equiv \{(\mathbf{U},\mathbf{V},\beta):\Vert (\mathbf{U},\mathbf{V},\beta) - (\bar{\mathbf{U}},\bar{\mathbf{V}},\bar{\beta})\Vert_F \le \rho\} $, i.e., there exists $ \theta \in [0,1) $ and $ M > 0 $ such that 
\begin{equation} \nonumber
\frac{\lvert F(\mathbf{U},\mathbf{V},\beta) - F(\bar{\mathbf{U}},\bar{\mathbf{V}},\bar{\beta})\rvert^\theta} {\text{dist}(\mathbf{0},\partial F(\mathbf{U},\mathbf{V},\beta))} \le M \quad \forall (\mathbf{U},\mathbf{V},\beta) \in \mathcal{B}_\rho(\bar{\mathbf{U}},\bar{\mathbf{V}},\bar{\beta}).
\end{equation} 

\noindent 
From \eqref{eq:Finequal}, note that $ F(\mathbf{U}^k,\mathbf{V}^k,\beta^k) $ is monotonically nonincreasing and also $ F(\mathbf{U}^k,\mathbf{V}^k,\beta^k) \ge F(\bar{\mathbf{U}},\bar{\mathbf{V}},\bar{\beta}) $  for all $ k $. Then, 
articles \cite{Shi2014,Xu2013} show that the KL inequality ensures that if we assume $ (\mathbf{U}^k,\mathbf{V}^k,\beta^k) \in \mathcal{B}_\rho(\bar{\mathbf{U}},\bar{\mathbf{V}},\bar{\beta}) $ for $ 0 \le k \le N $, then $ (\mathbf{U}^{N+1},\mathbf{V}^{N+1},\beta^{N+1}) \in \mathcal{B}_\rho(\bar{\mathbf{U}},\bar{\mathbf{V}},\bar{\beta}) $. Thus, $ (\mathbf{U}^k,\mathbf{V}^k,\beta^k) \in \mathcal{B}_\rho(\bar{\mathbf{U}},\bar{\mathbf{V}},\bar{\beta}) $ for all $ k $ by induction. In addition, they also show the following:
\begin{equation} \nonumber
\sum_{k=1}^{\infty} \Vert (\mathbf{U}^k,\mathbf{V}^k,\beta^k) - (\mathbf{U}^{k+1},\mathbf{V}^{k+1},\beta^{k+1}) \Vert_F < \infty,
\end{equation}
\noindent which indicates that $ \{(\mathbf{U}^k,\mathbf{V}^k,\beta^k)\} $ is a Cauchy sequence and thus converges to the limit point $ (\bar{\mathbf{U}},\bar{\mathbf{V}},\bar{\beta}) $. As a result, $ \{(\mathbf{U}^k,\mathbf{V}^k,\beta^k)\} $ converges to a critical point $ (\bar{\mathbf{U}},\bar{\mathbf{V}},\bar{\beta}) $ by Lemma C1. The detailed proof can be found in \cite{Shi2014,Xu2013}.

\setlength{\bibsep}{1pt}

\end{document}